\documentclass[a4paper]{article}
\usepackage{multirow}
\usepackage{amsmath}
\usepackage{indentfirst}
\usepackage{lscape}
\usepackage{booktabs}
\usepackage{graphicx,xcolor}
\begin{document}

\begin{titlepage}

\title{Laboratory demonstration of a cryogenic deformable mirror for wavefront correction of space-borne infrared telescopes}

%\begin{flushleft}
\author{Aoi Takahashi $^{1,2,3,*}$ \and Keigo Enya $^{2,1}$ \and Kanae Haze $^{4}$ \and Hirokazu Kataza $^{2,5}$ \and Takayuki Kotani $^{6}$ \and Hideo Matsuhara $^{2,1}$ \and Tomohiro Kamiya $^{7}$ \and Tomoyasu Yamamuro $^{8}$ \and Paul Bierden $^{9}$ \and Steven Cornelissen $^{9}$ \and Charlie Lam $^{9}$ \and Michael Feinberg $^{9}$}
%\end{flushleft}

\date{\today}
\maketitle

{\small {\sl
\begin{enumerate}
\item Department of Space and Astronautical Science, The Graduate University for Advanced Studies (SOKENDAI), Shonan Village, Hayama, Kanagawa 240-0193, Japan
\item Institute of Space and Astronautical Science, Japan Aerospace Exploration Agency, 3-1-1 Yoshinodai, Chuo-ku, Sagamihara, Kanagawa 252-5210, Japan
\item School of Science and Technology, Kwansei Gakuin University, 2-1 gakuen, Sanda, Hyogo 669-1337, Japan
\item Tokyo Office, Japan Aerospace Exploration Agency, Ochanomizu sola city, 4-6 Kandasurugadai, Chiyoda-ku, Tokyo 101-8008, Japan
\item Department of Astronomy, Graduate School of Science, The University of Tokyo, 7-3-1 Hongo, Bunkyo-ku, Tokyo 113-0033, Japan
\item National Astronomical Observatory of Japan, Extrasolar Planet Detection Project Office, 2-21-1 Osawa, Mitaka, Tokyo 181-8588, Japan
\item Tsukuba Space Center, Japan Aerospace Exploration Agency, 2-1-1 Sengen, Tsukuba, Ibaraki 305-8505, Japan
\item OptCraft, 3-16-8 Higashi-Hashimoto, Midori-ku, Sagamihara, Kanagawa, 252-0144, Japan
\item Boston Micromachines Corporation, 30 Spinelli Place, Cambridge, MA 02138, USA
\end{enumerate}
* Corresponding author: aoi@ir.isas.jaxa.jp
}}\\

%\thispagestyle{empty}
%\end{titlepage}

\begin{abstract}

This paper demonstrates a cryogenic deformable mirror (DM) with 1,020 actuators based on micro-electrical mechanical systems (MEMS) technology. Cryogenic space-borne infrared telescopes can experience a wavefront error due to a figure error of their mirror surface, which makes the imaging performance worse. 
For on-orbit wavefront correction as one solution, we developed a MEMS-processed electro-static DM with a special surrounding structure for use under the cryogenic temperature. We conducted a laboratory demonstration of its operation in three cooling cycles between 5 K and 295 K. Using a laser interferometer, we detected the deformation corresponding to the applied voltages under the cryogenic temperature for the first time. The relationship between voltages and displacements was qualitatively expressed by the quadratic function, which is assumed based on the principle of electro-static DMs. We also found that it had a high operating repeatability of a few nm RMS and no significant hysteresis. 
Using the measured values of repeatability, we simulated the improvement of PSF by wavefront correction with our DM. These results show that our developed DM is effective in improving imaging performance and PSF contrast of space-borne infrared telescopes.
\end{abstract}
\end{titlepage}
%\begin{document}

\maketitle
%\thispagestyle{fancy}

%\ifthenelse{\boolean{shortarticle}}{\ifthenelse{\boolean{singlecolumn}}{\abscontentformatted}{\abscontent}}{}

\section{Introduction}

\subsection{Wavefront error in space-borne infrared telescopes}
Space-borne infrared telescopes are essential for mid-to-far-infrared observations, because they are not constrained by atmospheric windows. They are also removed from the thermal radiation in the atmosphere and can detect much fainter objects than ground-based telescopes with the same aperture. Normally, they are cooled to the cryogenic temperature by liquid helium and/or mechanical coolers to reduce their own thermal radiation. Table \ref{telescope} indicates examples of their cooling temperatures.

\begin{table}[t]
\centering
\caption{\bf Specifications of space-borne infrared telescopes. ($\ast$ in Launch year means just planning.)}
%\small
\scalebox{0.8}{
\begin{tabular}{lcccc} \toprule
            Telescope & Launch [year] & Apperture [m] & Wavelength [$\mu$m] & Cooling temperature [K] \\ \midrule
            {\sl IRAS}~\cite{telescope1} & 1983 & 0.57 & 10 -- 100 & $<3$ \\
            {\sl IRTS}~\cite{telescope2} & 1995 & 0.15 & 1.4 -- 700 & 1.9 \\
            {\sl ISO}~\cite{telescope3} & 1995 & 0.6 & 2.5 -- 240 & 3 \\
            {\sl Spitzer Space Telescope}~\cite{telescope4} & 2003 & 0.85 & 3.6 -- 160 & 5.5 \\
            {\sl AKARI}~\cite{telescope5} & 2006 & 0.68 & 2 -- 180 & 5.8 \\
            {\sl WISE}~\cite{telescope6} & 2009 & 0.4 & 3.3 -- 23 & 17 \\
            {\sl Herschel Space Observatory}~\cite{telescope7} & 2009 & 3.5 & 55 -- 671 & $ \simeq 85$ \\
            {\sl JWST}~\cite{telescope8} & $2018^{\ast}$ & 6.5 & 0.6 -- 28 & $ \simeq 40$ \\
            {\sl SPICA}~\cite{telescope9} & 2027 -- $28^{\ast}$ & 2.5 & 12 -- 210 & 8 \\ \bottomrule
   \end{tabular}
   }
  \label{telescope}
\end{table}

Although such telescopes are immune to air perturbations, they can experience a wavefront error. In addition to a figure error of the mirror surface caused by the manufacturing process, the mirror can be deformed by release from gravity and the cooling of the whole telescope mentioned above. Light reflected by such a deformed surface would include a wavefront error. This error can degrade the imaging performance.

Direct observation of exoplanets is one of the observations most sensitive to a wavefront error. A wide dynamic range is needed for this observation because planets are very faint as compared with the central star. The earth, for example, is $10^{-10}$ times fainter than the sun in the visible region, and $10^{-6}$ times so in the mid-infrared region if we look at our solar system from outside~\cite{Solar_system_contrast}.
From this reason, we generally use coronagraph optics that control the point spread functions (PSFs) of central stars and form "dark regions" where the surrounding diffracted light is suppressed. This allows us to detect faint signals from exoplanets in the dark region. In combination with coronagraphs, space-borne infrared telescopes enable direct observation of faint exoplanets that ground-based telescopes cannot detect. However, such observations using coronagraphs are sensitive to wavefront error of the incident light. Scattered light from central stars (speckle) contaminates dark regions and degrades the PSF contrast in there. Faint exoplanets cannot be detected if they are buried in those speckles.

\subsection{Wavefront correction by a cryogenic deformable mirror (DM)}

One solution is wavefront correction on orbit, which we plan to do using a small, deformable mirror (DM) added to the back-end optics of space-borne infrared telescopes. For this purpose, we need a thin DM with many actuators that is usable under the cryogenic temperatures typical of cooling in space-borne infrared telescopes. No conventional DM, however, satisfies such requirements (see section \ref{DM} for detail). We are therefore developing a micro-electrical mechanical systems (MEMS)-processed electro-static DM usable under the cryogenic temperatures. The prototype with 32 actuators has already been demonstrated to operate at 95 K~\cite{95K_DM}. We demonstrated and evaluated the operation of a newly developed DM with 1,020 actuators at 5 K. This paper represents the results of the DM operations measured over three cooling cycles between 5 K and 295 K.

\subsection{Flow of this paper}

This paper is structured as follows. In the 2nd section, we show the detailed specification and structure of our developed DM. The 3rd section indicates the procedure for the cooling tests, and the 4th section presents the results and the simple discussion. Using the results, we simulated a wavefront corrected by the DM, and the expected astronomical outcome in the 5th section. Finally, the 6th section presents our summary.

\section{Developed cryogenic deformable mirror}
\label{DM}

\subsection{Principle and specifications}
The DM must have the following properties for our purposes. At first, it should deform as intended under the cryogenic temperatures where space-borne infrared telescopes are generally cooled. Secondly, multiple actuators are also necessary to correct a wavefront error with typical spatial frequency for these telescopes. Furthermore, it can use only strictly limited resources of space and electric power. On the other hand, these telescopes do not have a wavefront error caused by the perturbations in the air. Therefore, it does not need speedy response, since the wavefront error does not change over an observation time scale.

\begin{figure}[tb]
\centering
\includegraphics[width=0.7\linewidth]{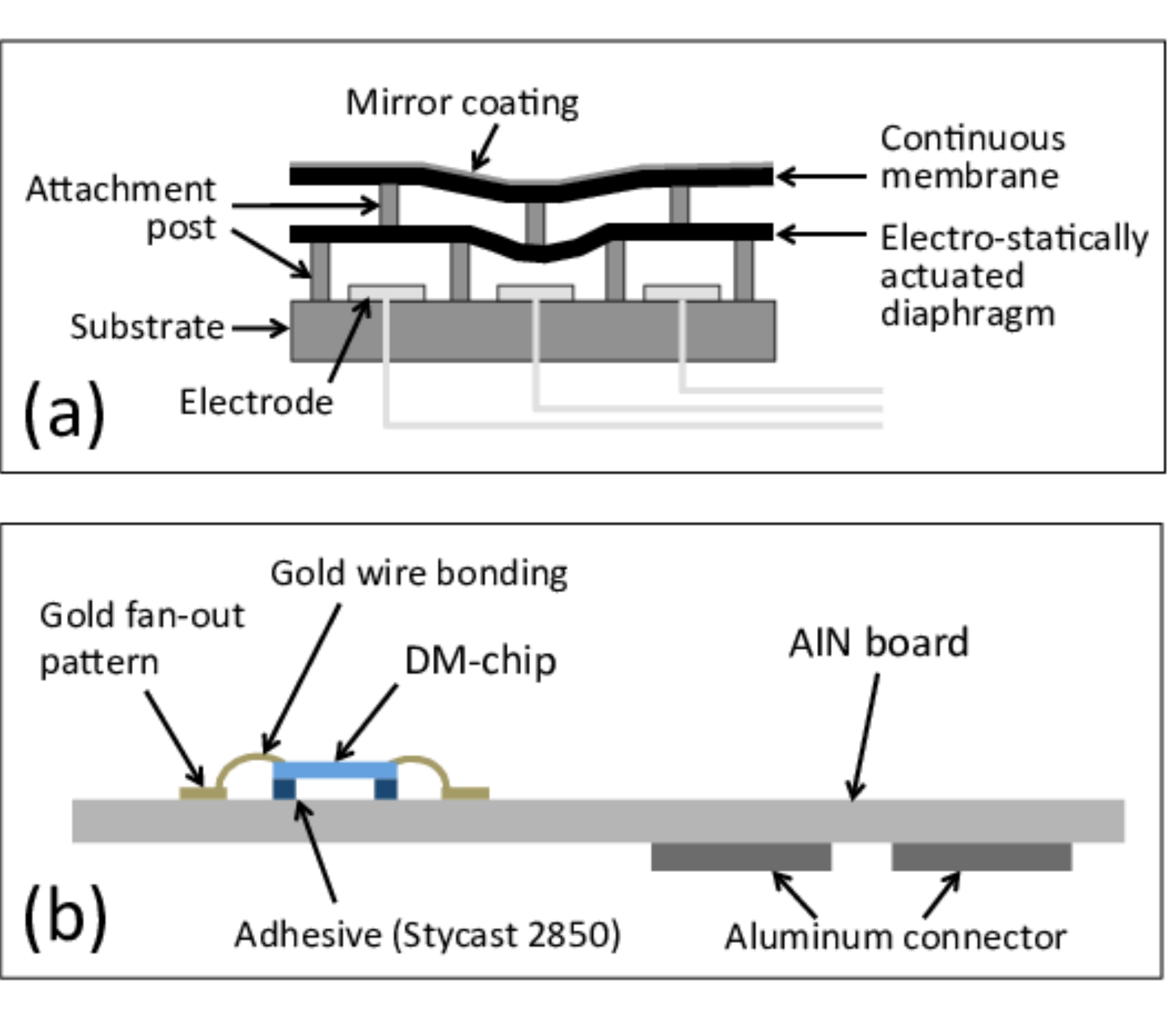}
\caption{The cross-sectional schematic view of (a) DM chip and (b) a new DM unit.}
\label{DM_shem}
\end{figure}

We reviewed types of DM based on these requirements. DMs using electro-magnetic force need a magnet and a coil for each actuator in the backside of mirror surface and are not thin enough to be easily contained in the limited space. The piezoelectric effect, which is often used as the driving principle of DM, is known to be much weaker and cause much less deformation under cryogenic temperatures than under room temperatures~\cite{piezo}. In contrast, electro-static DMs can not only easily realize a DM with multi actuators in compact depth, but are also expected to deform independently of the temperature due to being driven by Coulomb forces. In addition, little electric power is needed for driving. This type of DM with double layer structure and membrane surface does not worsen optical performance, unlike DMs with segmented surfaces. For these reasons, we adopt a MEMS-processed electro-static type of DM with a double-layered structure and membrane surface. The cross-sectional schematic view is shown in Figure \ref{DM_shem} (a).

We used a DM chip conventionally manufactured by Boston Micromachines Corporation (BMC), with 32 $\times$ 32 actuators arranged in a square, although the four corner actuators cannot be displaced due to being fixed. We can deform the mirror surface by applying a voltage to each of the 1,020 unfixed actuators. This means that we can exclude speckles in a region up to 16 $\lambda/D$ [rad] from a PSF peak, if $\lambda$ and $D$ mean the observation wavelength and the aperture diameter, respectively. The interval of the actuators is 300 $\mu$m and the maximum stroke of each actuator is 1.5 $\mu$m, according to the corporation. This stroke is enough to correct the wavefront error assumed by SPICA in 2012 of 350 nm RMS ( $\sim$ 2700 nm PV )~\cite{SPICA_WFE}, for example. The mirror size of 9.6 mm $\times$ 9.6 mm and the thickness of a few mm enable insertion into the limited space for optics. The mirror surface is coated in gold, which has a reflectance above 95 $\%$ in infrared. 

\subsection{Unit structure}
\label{DM_unit}
\begin{figure}[tb]
\centering
\includegraphics[width=0.7\linewidth]{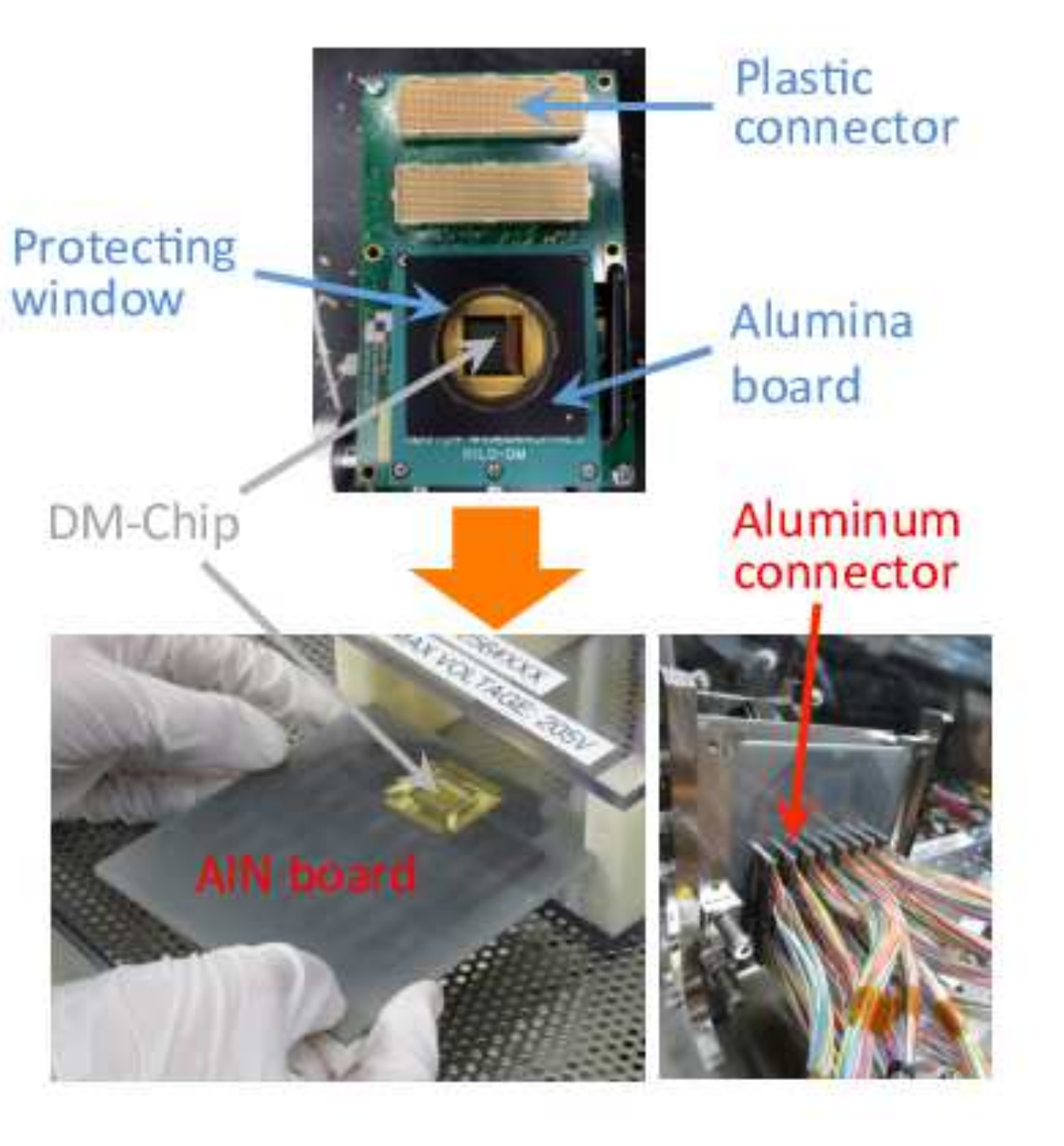}
\caption{Picture of a conventional DM unit (Top) and our developed cryogenic DM unit (front side view in left bottom, back side view in right bottom). The parts indicated in blue experience problems at cryogenic temperatures, so we changed them to new parts indicated in red.}
\label{DM_pic}
\end{figure}

Conventional MEMS-processed DMs are, however, inadequate for use under cryogenic temperatures from the point of their whole-unit structure. Therefore, we developed a new DM unit (Figure \ref{DM_shem} (b)). In conventional DMs, a DM chip made of silicon and the supporting-board made of alumina have different coefficients of thermal expansion (CTE), causing thermal stress when cooling. To reduce stress, we glued a DM chip onto a circuit board made of aluminum nitride (AlN) because it has a similar CTE to that of silicon and a thermal conductivity 5-7-times higher than alumina. We used the epoxy adhesive "STYCAST 2850FT" for cryogenic use. The DM chip is glued at four corners such that not only does its adhesive area become small but also stable gluing is maintained, even if one point of glue peels off. We also changed the type of connector attached to the circuit board from a conventional plastic one to aluminum one tolerant of cryogenic environments. Finally, conventional DMs have a protecting window made of glass that is sealed with adhesive and filled with nitrogen gas. We did not install this window because it could be damaged due to stress in vacuum drawing or cooling. These changes are shown in Figure \ref{DM_pic}. 

Work sharing between ISAS and BMC was organized as follows. First, ISAS team presented conceptual design and determined the gluing process of a DM chip from the results of tests using gluing samples made of silicon. In the next step, more realistic samples, but without electrical connection, were manufactured in BMC, and the cryogenic tests were carried out in ISAS. BMC also completed the thermal analysis. Based on these results, design and manufacturing process were fixed and our DM was manufactured in BMC. After pre-shipment tests at room temperature in the air in BMC, the DM was finally delivered to ISAS, and the test presented in this paper were pursued in ISAS.

\section{Cooling tests}
\label{Cooling tests}

\subsection{Purpose and overview}
In this test, we cooled our developed DM to 5 K, applied voltages to each actuator, and measured the surface figure. Our primary purpose was to demonstrate the operation of the DM at cryogenic temperature. As the second purpose, we aimed to characterize the relationship between the applied voltages and the displacement of the actuators (called "operating characteristic (OC)" in this paper), and derive the hysteresis and repeatability. The displacement resolution of the measurement system was also checked to distinguish the real response of the DM from error originating in the measurement system. In addition, we tested the durability over three cooling cycles between 295 K (normal phase) and 5 K (cryogenic phase).  In each phase, we measured the surface figure with no applied voltage, the OC, the repeatability of our DM and the displacement resolution of the measurement system.

\subsection{Test system}
\begin{figure}[tb]
\centering
\includegraphics[height=\linewidth, angle=-90]{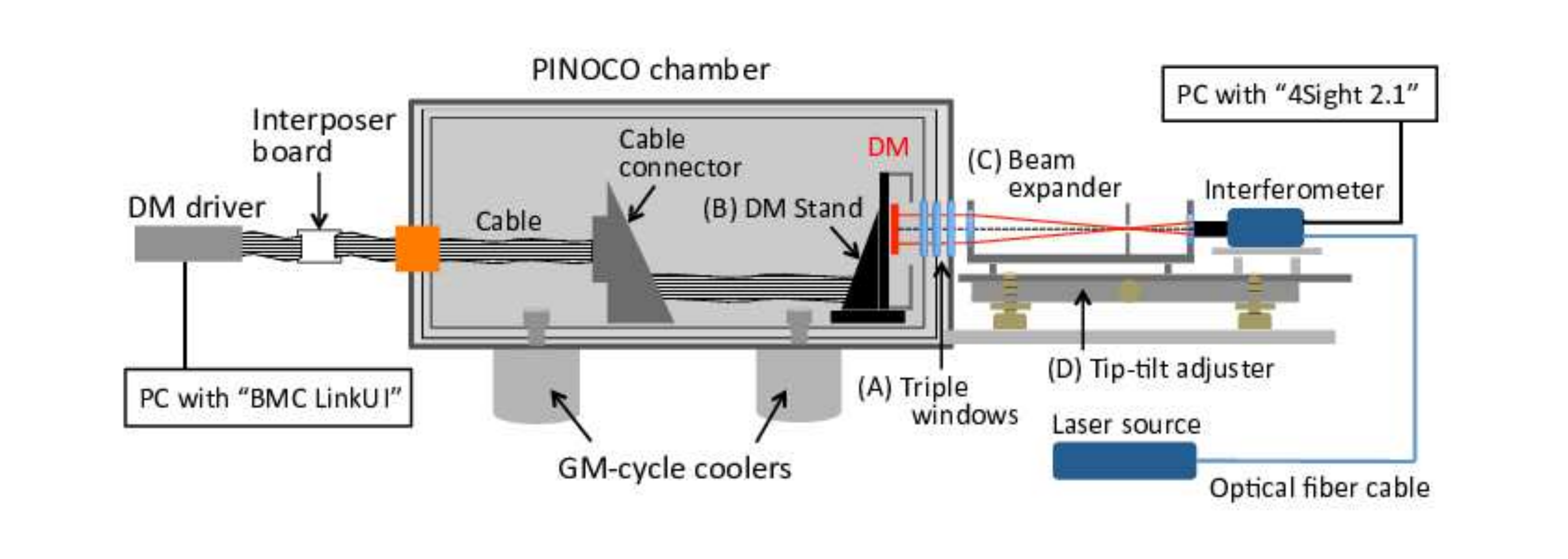}
\caption{Assembled test system}
\label{experiment_system}
\end{figure}

\begin{figure}[tb]
\centering
\includegraphics[height=\linewidth, angle=-90, scale=0.8]{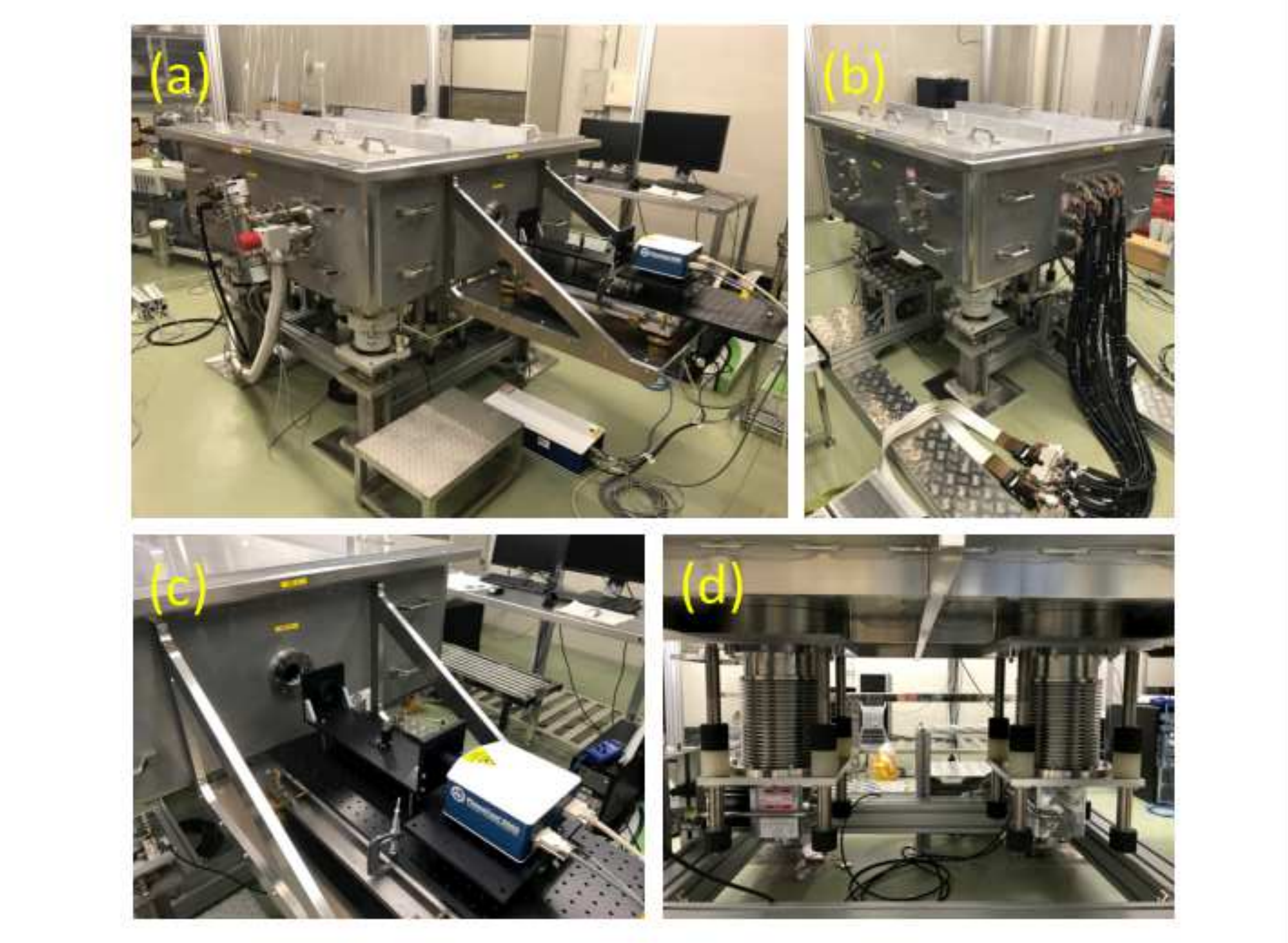}
\caption{Picture of test system: (a) window-side view of PINOCO; (b) driver-side view of PINOCO; (c) measurement system in front of the window; (d) GM-cycle coolers}
\label{pinoco_pic}
\end{figure}

For cooling tests, we set up the test system as in Figures \ref{experiment_system}, \ref{pinoco_pic}.
It included the following equipment. 

\subsubsection{Cooling system}
We used the chamber "PINOCO"~\cite{PINOCO} to cool the DM. The optical bench was covered with a double radiation shield of aluminum and multiple layers of insulation. GM-cycle mechanical coolers with two stages cooled the optical bench to 5 K using only electric power. They are attached to the bottom of the chamber through soft structures including the damper. Additionally, the cooler heads are connected to the optical bench through braided copper wires. They can prevent transmission of vibrations from coolers to the chamber.

We can measure temperature in the chamber using eight silicon diode sensors. In these tests, we pasted each sensor to each position including the optical bench (channel A), the leg of the DM stand (channel B), and the backside of the DM's AlN board (channel C), and automatically recorded their temperatures each minute. The temperature of the DM in each phase of the cooling cycle was defined by the equilibrium temperature of channels B and C. Cooling took about a week and warming took about 10 days. In the cryogenic phases of each cooling cycle, channel B and C indicated 6.1 K - 6.2 K and 5.0 K - 5.1 K, respectively. About normal phases, the temperature differed by a few K depending on cooling cycles due to difference in the atmospheric temperature. In spite of this, It was set in range from 294.5 K to 297.5 K in all cooling cycles. We also note that channel B was about 1.2 K lower than channel C in normal phases of each cooling cycle, in contrast to cryogenic phases. The channel-dependent difference seems to be an individual difference of sensors.

PINOCO has the following functions other from cooling. At the side of the chamber and radiation shields, there are triple windows with 6-cm diameter (see (A) in Figure \ref{experiment_system}). These are made of BK7 glass and are coated anti-reflectively in broad optical band. While they are highly transparent to optical light, they cut off infrared and protect the chamber against thermal inflow through the window. The optical laser light passes through these windows in this tests. In front of the windows, a stage was set for additional optics such as interferometers. At the other side of the chamber, there are 16 connectors for cables from the DM in the chamber to the DM control device outside of the chamber.

For this test, we designed and constructed a DM stand to hold the mirror on optical bench (see (B) in Figure \ref{experiment_system}). In our design, an AlN board is first inserted into the cavity of the stand with slight margins in all directions. Secondly, we softly fix the AlN board from the backside using spring screws to reduce thermal stress to the AlN board during cooling cycles. The stand is made of oxygen-free copper (C1020), same as the optical bench in PINOCO. Note that we designed the stand such that the height of the DM chip matches that of the chamber window's center.

\subsubsection{Measurement system}
We used the PhaseCam 6000 dynamic laser interferometer (4D Technology Company) to measure the mirror surface figure. The measurements are performed through special software, "4Sight 2.1", installed on a PC. The interferometer can measure wavefront in 30 $\mu$ seconds, 1/5,000 of the time required by a phase-shift laser interferometer. This reduces vibration effects in measurements, allowing us to obtain high stability. This interferometer radiated a collimated He-Ne light of diameter 9.0 mm and wavelength 632.8 nm. Because the diameter was insufficient to cover the whole mirror surface, we designed and manufactured a beam expander (see (C) in Figure \ref{experiment_system}). Its magnification is about 1.67, expanding the beam's diameter to 15.0 mm. We designed it to horizontally align the light axis from the interferometer to the center of the chamber window. Furthermore, it was attached to a breadboard by three points.

Due to thermal deformation and vibration, the DM's tilt and position against the laser-light axis slightly change during cooling cycles. Therefore, we established a system to adjust the tip-tilt and position of the laser-light axis. As you can see in Figure \ref{experiment_system} (D), we put the interferometer and the beam expander on the same breadboard, and then adjusted its tip-tilt and position as the whole breadboard. We identified the DM's tilt against the laser-light axis using the pinhole board in the focal plane of the beam expander. As a requirement of this adjustment, the focal point of the light reflected by DM should match the position of the pinhole in the beam expander. For a more detailed adjustment, we looked at the images of the interference fringes and slightly changed the tip-tilt and position of the laser-light axis such that the fringes became widest. We performed this adjustment before the first measurement in every phase. 

\subsubsection{Control system}
\label{control_system}
To deform and control the DM's mirror surface, we applied voltages to actuators through a DM driver and a PC on which the special software "BMC LinkUI" was installed. In case of this DM driver, we input a 16-bit value as 4 characters of hexadecimal number for each actuator, and then the corresponding voltages were applied to the actuators. Since the max value of 16-bit is 65535, we can apply up to 285 V to each actuator with a resolution of 285 V / 65535 $\sim$ 4 mV. The frame rate was more than 30 kHz in ambient conditions. Although we can control the value, the maximum frame rate is dependent on a number of factors in the controlling system.

\begin{figure}[tb]
\centering
\includegraphics[height=\linewidth, angle=-90, scale=0.8]{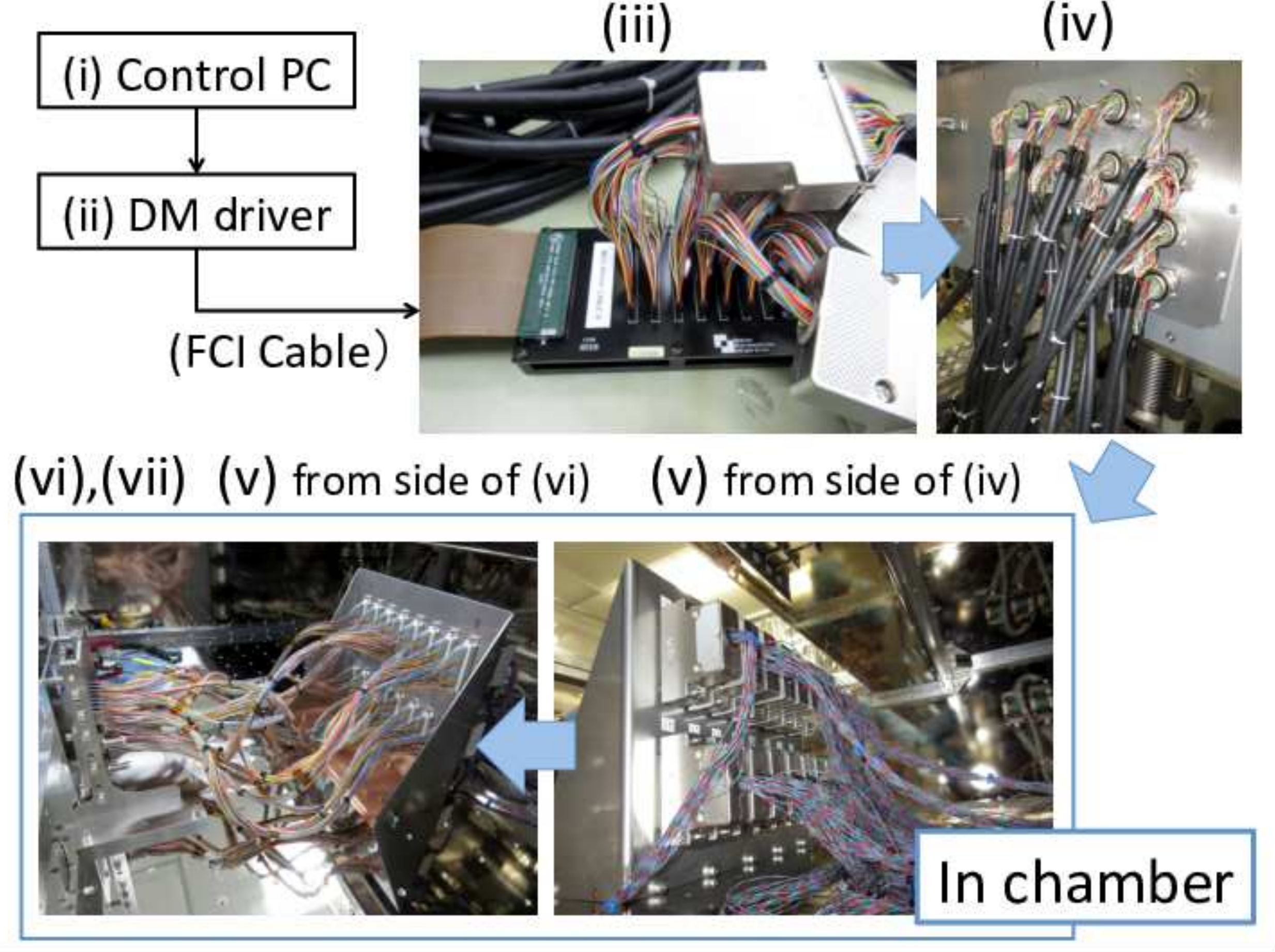}
\caption{Pathway of the wiring system for DM control: (i) Control PC; (ii) DM driver; (iii) Interposer board; (iv) Side of PINOCO chamber; (v) Cable Connector; (vi) Back side of AlN board; (vii) DM actuators}
\label{haisen}
\end{figure}

We prepared a special wiring system to connect the DM in the chamber to the driver outside of the chamber. The pathway of this system is shown in Figure \ref{haisen}. Because of parallel controlling, we need 1,056 channels in total: 1,020 for actuators and 36 for grounds. To reduce thermal inflow through the wiring, we used manganic thin wires with low thermal conductivity in the chamber. Outside of the chamber, copper wires were used.

\subsection{Content of the measurements}

\subsubsection{Voltage distribution}
\begin{figure}[htbp]
\centering
\includegraphics[width=0.6\linewidth]{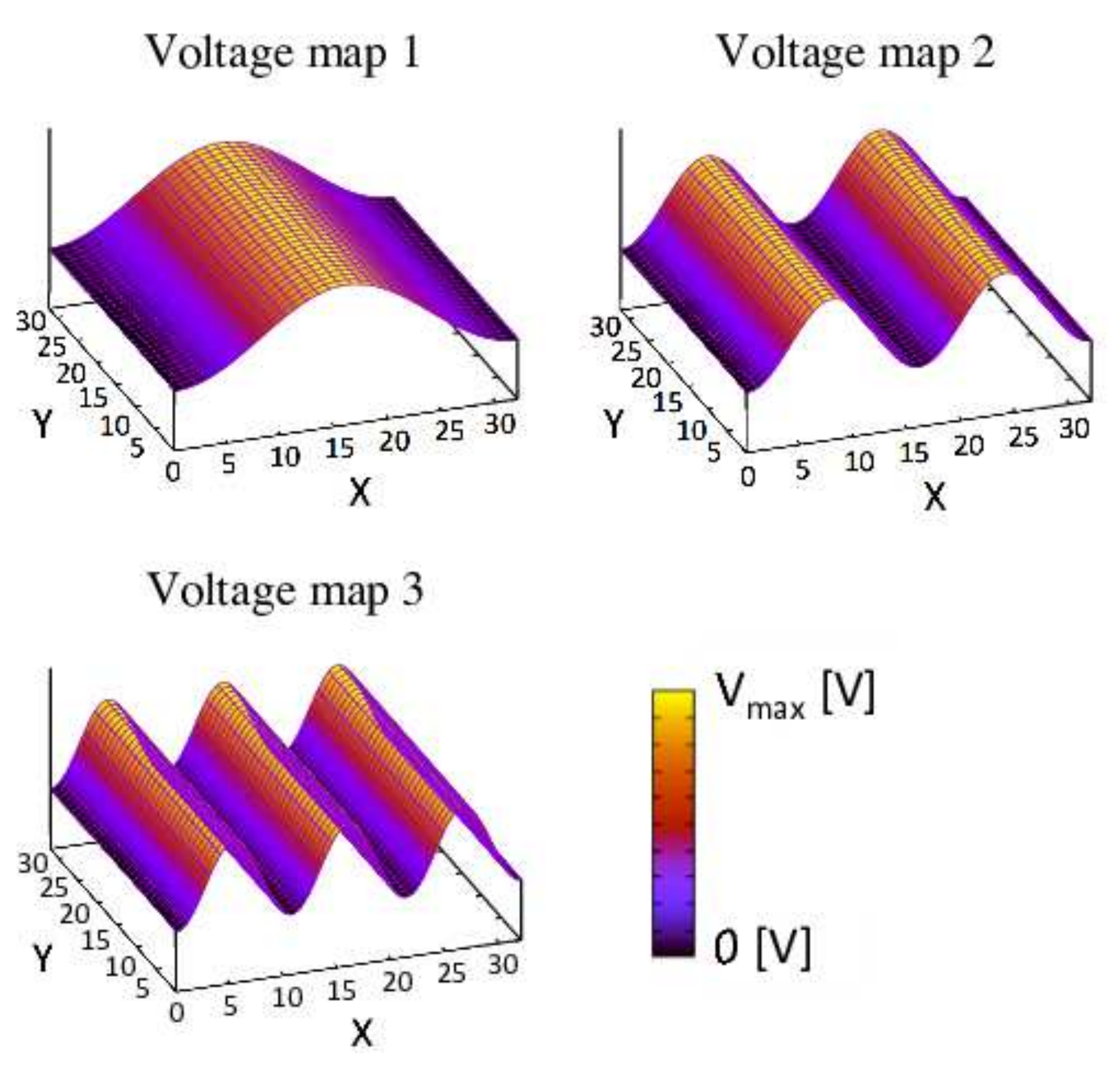}
\caption{Applied voltage distributions}
\label{voltage_map}
\end{figure}

In this test, we applied three types of voltage distribution to the mirror surface. These are expressed as sinusoidal waves in one dimension. We call the Voltage distribution with spatial frequency of N cycle/D [m$^{-1}$] "Voltage map N", for N = 1, 2, 3. They are written by the following equation:
\begin{eqnarray}
V(x,y) = \frac{V_{max}}{2} \left( \sin \left( 2N \pi x - \frac{\pi}{2} \right) + 1 \right) \mbox{~~~~for N = 1, 2, 3}
\end{eqnarray}
D indicates a side length of the DM surface, 9.6 mm. We defined the (x,y) plane such that the regions of $0 \leq x \leq 1$ and $0 \leq y \leq 1$ fitted on the DM's surface. $V_{max}$ means the maximum voltage we applied, which corresponds to double of the sinusoidal amplitude. Figure \ref{voltage_map} shows these distributions. We used five voltage distributions with $V_{max}$ of 20, 40, 60, 80, and 100 V for each type of voltage map. A voltage distribution was input to the DM-controller software as a text file with 1,024 hexadecimal numbers and applied to the actuators, including four fixed actuators at corners.

\subsubsection{Data processing}
\label{data_process}
\begin{figure}[htbp]
\centering
\includegraphics[width=0.5\linewidth]{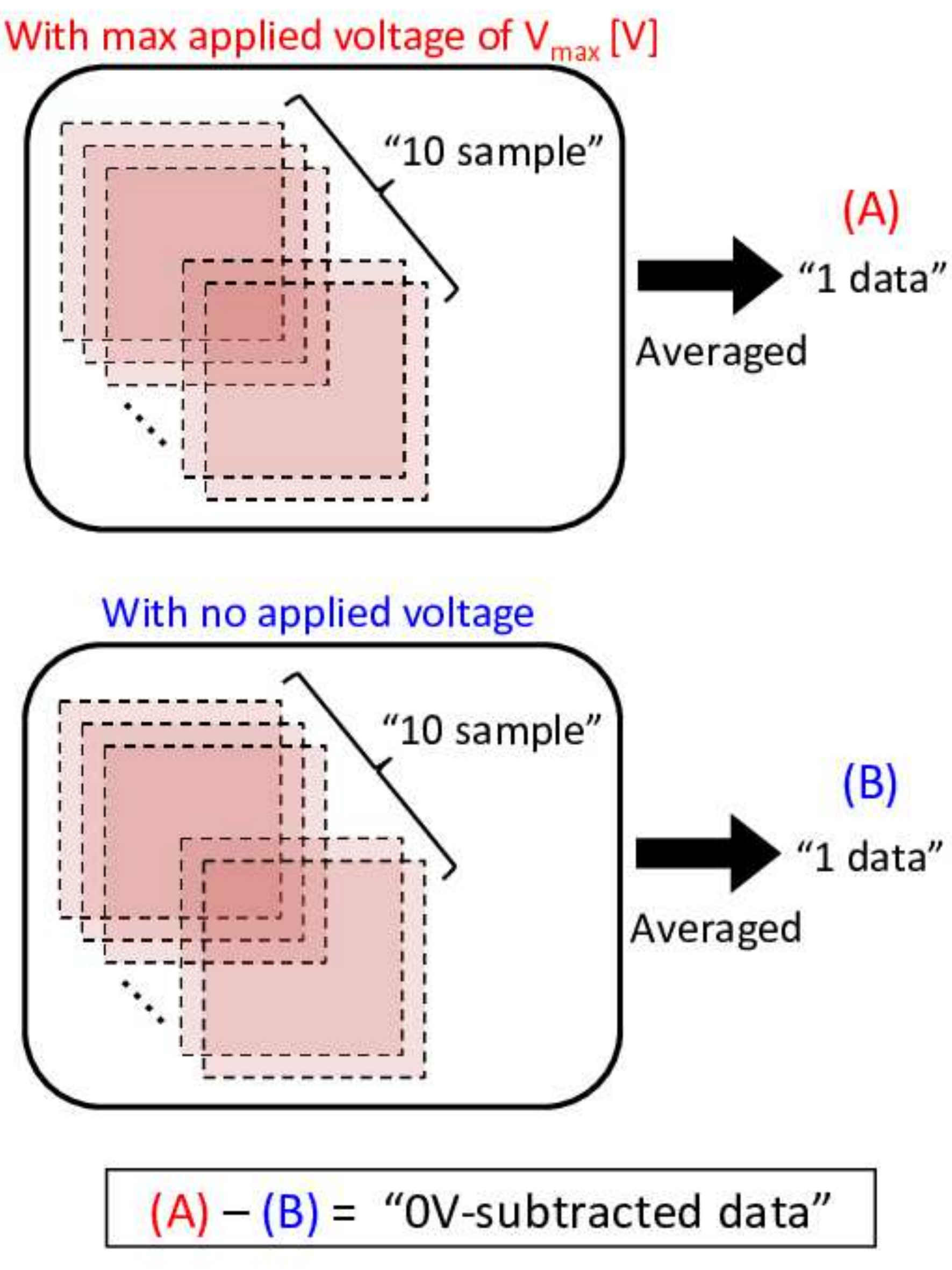}
\caption{Flow of data processing}
\label{data_name}
\end{figure}

We took ten data samples and their average was outputted as measurement data. To estimate displacements of actuators due to voltage-applications, measurement data with no applied voltage was subtracted from those with applied voltages. We defined data after this subtraction as "0V-subtracted data", and Figure \ref{data_name} indicates this relations.

We need software-masks that extract the region of DM's surface from the field of view. The design process was as follows. First, in normal phases, we measured the surface figure under application of 60 V to edge actuators on the DM's surface. The peak position of concavity indicates the central position of each edge actuator; therefore, we can calculate the edge position of DM's surface by half-actuator extension to outer direction. We then fitted the edge of the software-mask to the edge position of the DM's surface and extracted data only in the mask region. In cryogenic phases, however, it was difficult to measure the surface figure on the edge of DM's surface, because it included lack or discrete values (see section \ref{Result}.\ref{no applied-voltages}, \ref{Result}.\ref{deformation} for more detail). Instead of applying voltage to edge actuators, we applied it to actuators in specific rows and columns, and calculated the edge position of the DM's surface from the data in the cryogenic phases. We made a software-mask for each phase of the cooling cycles and extracted the data in the region of the DM's surface from the 0V-subtracted datasets.

\subsubsection{Measurement item}
\begin{description}
\item[(i) A measurement with no applied voltage] \mbox{}\\
To estimate the deformation due to vacuum drawing and/or change in temperature including permanent deformation, we measured the surface figure with no applied voltage initially in each phase.
\item[(ii) Five measurements of the same surface figure] \mbox{}\\
To estimate the resolution of the measurement system, we measured the same surface figure continuously five times keeping the timely constant voltages applied. The interval time was about a few 10s seconds, because we assumed that the resolution was determined by un-stability of measurement systems due to high frequency vibration. In each phase, we conducted these measurements using Voltage map 1 with $V_{max}$ of 0, 20, 40, 60, 80, and 100 V.
\item[(iii) Measurements increasing and decreasing $V_{max}$] \mbox{}\\
To estimate the OC and hysteresis, we took a measurement data for each $V_{max}$, increasing and decreasing $V_{max}$ as 0, 20, 40, 60, 80, 100, 80, 60, 40, 20, 0 V in order. For estimation of the operating repeatability, we also repeated these measurement's set five times except in the initial phase at (295 K, 1 atm), where we measured the surface figure only in the 1st $V_{max}$-increase using Voltage map 1, 2, and 3. In cryogenic phase of the 1st cooling cycle, measurements in the 1st to 5th $V_{max}$-increase and decrease were completed using Voltage map 1, 2, and 3. In other phases, we measured them using only Voltage map 1.
\end{description}

These measurement items are summarized along time series in Table \ref{experiment_shedule}. 

\begin{landscape}
\begin{table}[htbp]
      \caption{Measurement items in the cooling tests}
      \label{experiment_shedule}
%      \footnotesize
      \begin{center}
\scalebox{0.8}{
            \begin{tabular}{|c|l|l|l|} \hline
            Cooling cycle & 295 K , 1 atm & 295 K , 0 atm & 5 K , 0 atm \\ \hline\hline
              \multirow{3}{*}{Initial} & (i) 1 data with no applied voltage & & \\
                                       & (ii) 5 data of the same surface with various $V_{max}$ & & \\
                                       & (iii) data in 1st $V_{max}$-increase with Voltage map 1,2,3 & & \\ \hline
            \multirow{6}{*}{1st cycle} & & (i) 1 data with no applied voltage & \\
                                       & & (ii) 5 data of the same surface with various $V_{max}$ & \\
                                       & & \shortstack{(iii) data in 1st to 5th $V_{max}$-increase and decrease \\ with Voltage map 1} & \\ \cline{2-4}
                                       & & & (i) 1 data with no applied voltage  \\
                                       & & & (ii) 5 data of the same surface with various $V_{max}$ \\
                                       & & & \shortstack{(iii) data in 1st to 5th $V_{max}$-increase and decrease \\ with Voltage map 1,2,3 }\\ \hline
            \multirow{6}{*}{2nd cycle} & & (i) 1 data with no applied voltage & \\
                                       & & (ii) 5 data of the same surface with various $V_{max}$ & \\
                                       & & \shortstack{(iii) data in 1st to 5th $V_{max}$-increase and decrease \\ with Voltage map 1} & \\ \cline{2-4}
                                       & & & (i) 1 data with no applied voltage \\
                                       & & & (ii) 5 data of the same surface with various $V_{max}$ \\
                                       & & & \shortstack{(iii) data in 1st to 5th $V_{max}$-increase and decrease \\ with Voltage map 1} \\ \hline
            \multirow{6}{*}{3rd cycle} & & (i) 1 data with no applied voltage & \\
                                       & & (ii) 5 data of the same surface with various $V_{max}$ & \\
                                       & & \shortstack{(iii) data in 1st to 5th $V_{max}$-increase and decrease \\ with Voltage map 1} & \\ \cline{2-4}
                                       & & & (i) 1 data with no applied voltage \\
                                       & & & (ii) 5 data of the same surface with various $V_{max}$ \\
                                       & & & \shortstack{(iii) data in 1st to 5th $V_{max}$-increase and decrease \\ with Voltage map 1} \\ \hline
                \multirow{6}{*}{Final} & & (i) 1 data with no applied voltage & \\
                                       & & (ii) 5 data of the same surface with various $V_{max}$ & \\
                                       & & \shortstack{(iii) data in 1st to 5th $V_{max}$-increase and decrease \\ with Voltage map 1} & \\ \cline{2-4}
                                       & (i) 1 data with no applied voltage & & \\
                                       & (ii) 5 data of the same surface with various $V_{max}$ & & \\
                                       & \shortstack{(iii) data in 1st to 5th $V_{max}$-increase and decrease \\ with Voltage map 1} & & \\ \hline
            \end{tabular}
            }
       \end{center}
\end{table}
\end{landscape}

\section{Result}
\label{Result}

\subsection{Surface figures with no applied voltage}
\label{no applied-voltages}
\begin{figure}[htbp]
\centering
\includegraphics[width=0.7\linewidth]{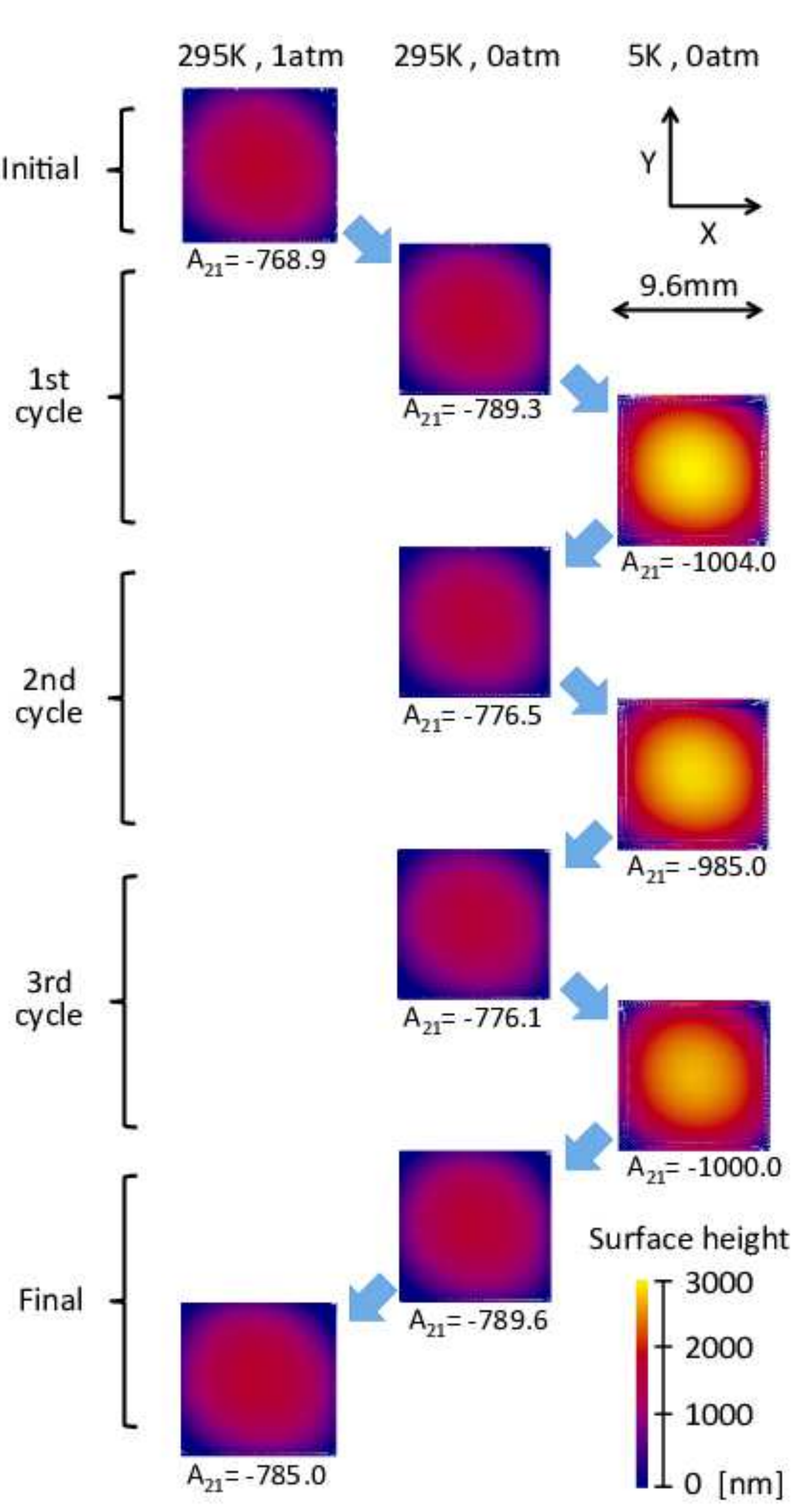}
\caption{Surface figures with no applied voltage in all phases of cooling cycles. Each square corresponds to a region of the DM's surface. Surface height of each figure was offset to zero at the minimum height, because we measured them as relative value in each figure. We corrected the tilt component caused by the laser-incidence angle in each figure. The gray-colored region indicates a lack of data.}
\label{color_map_0V}
\end{figure}

Destruction or clear permanent deformation of the DM's surface was not seen during cooling cycles. Figure \ref{color_map_0V} shows color maps of surface figures with no applied voltage in each phase. We calculated the Zernike decomposition for each figure using the whole square region and compared the amplitude of the defocus term, $A_{21}$ written under each square. This term means a component of concavity. According to Figure \ref{color_map_0V}, we obtain $A_{21} < 0$ and therefore the DM's surface figure is convex in all phases. Furthermore, the convexity is larger at 5 K than at 295 K. This can be qualitatively explained by the slight difference in CTE between the DM chip and the AlN board. The AlN board has a larger CTE than the DM chip made of silicon~\cite{Si}~\cite{AlN}, meaning that the AlN board contracted relative to the DM chip during cooling, pulling on the chip's four adhered corners. This can make the convexity of the DM's surface larger at 5 K.

In Figure \ref{color_map_0V} we can see the region with lack values and discreteness near the edges of the DM's surface. If we assume the case of relative contraction described above and the DM chip is pulled by the AlN board at the corners, the chip could be folded at the width of the adhered area from the edges. As a result, the DM's surface in the edge region could have an un-smooth figure and not be measured with certainty. 

\subsection{Measured surface deformation}
\label{deformation}

\begin{figure}[p]
\centering
\includegraphics[width=14cm]{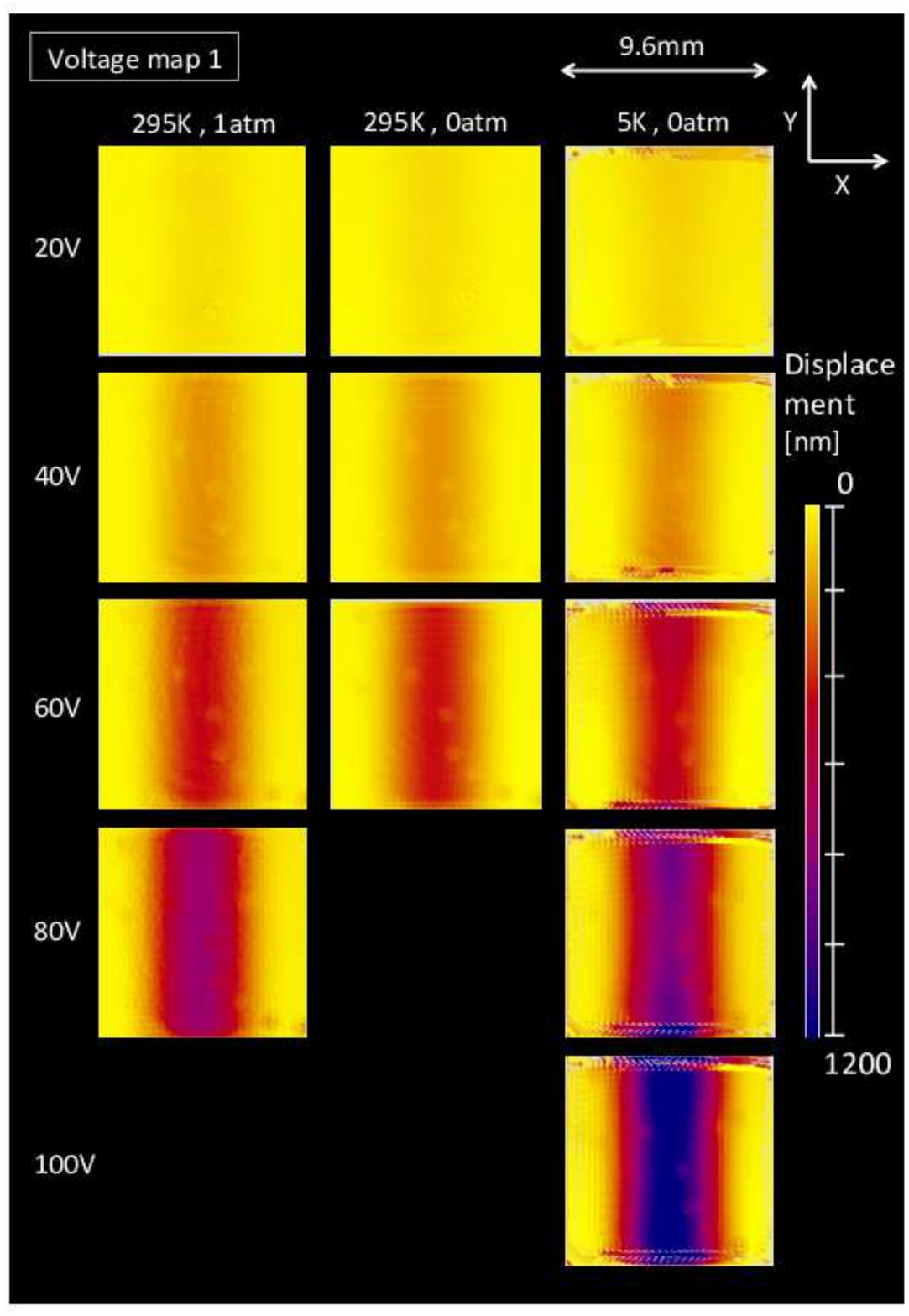}
\caption{The color maps of 0V-subtracted data in the case of Voltage map 1 with various $V_{max}$ in (295 K, 1 atm), (295 K, 0 atm), and (5 K, 0 atm) starting from the left column. Color maps in the same row show surface figures to which the same $V_{max}$ written in left of the row is applied. Each square corresponds to a region of the DM's surface. These measurements were taken in 1st $V_{max}$-increase during the initial phase and the 1st cycle. Displacement of each figure was offset to zero in edge region, because we measured their surface height as relative value in each figure. The gray-colored region indicates a lack of data.}
\label{color_map_vm1}
\end{figure}

\begin{figure}[p]
\centering
\includegraphics[width=14cm]{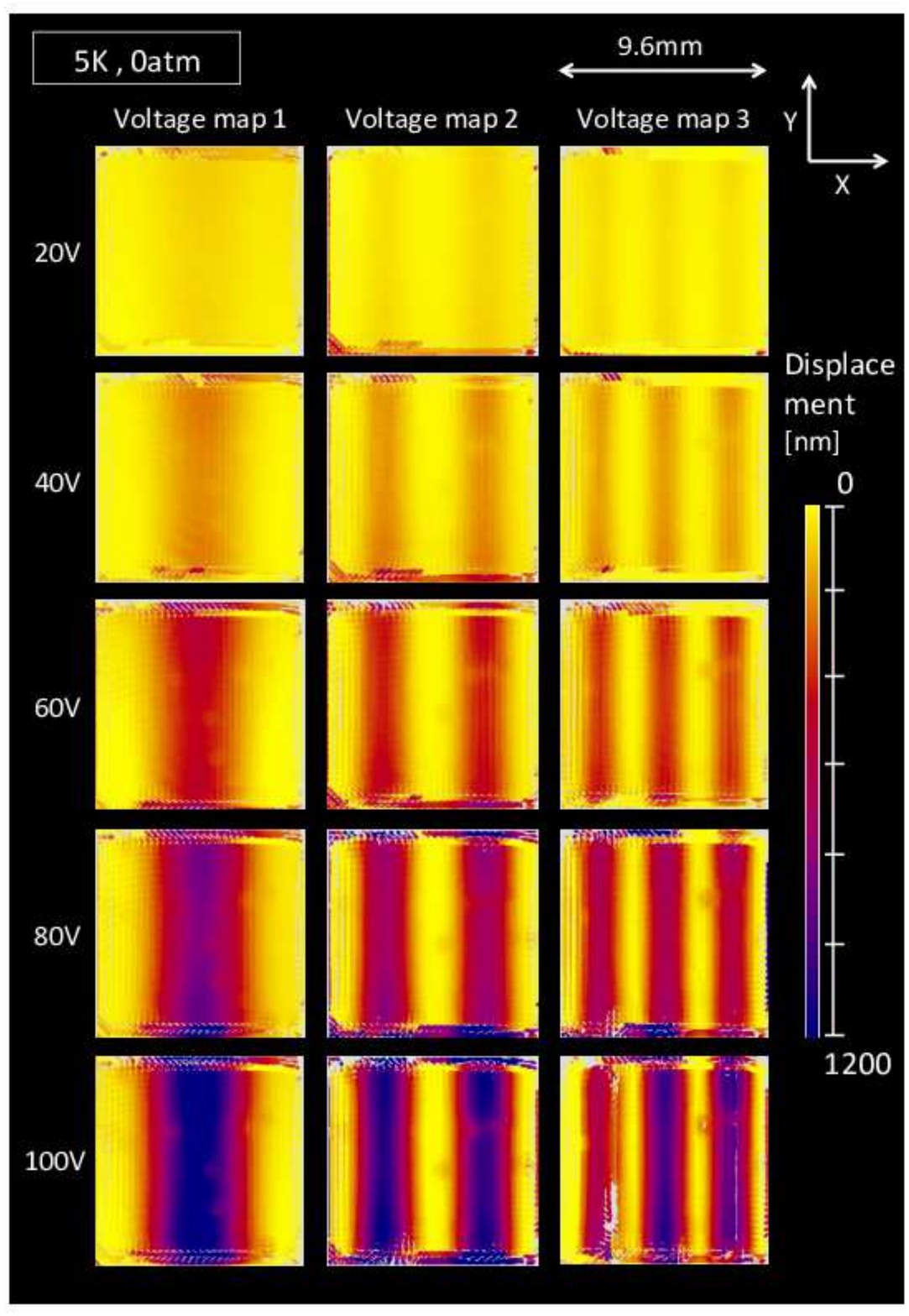}
\caption{The color maps of 0V-subtracted data for Voltage map 1, 2, and 3 with various $V_{max}$ values in (5 K, 0 atm). The format of this figure is the same as Figure \ref{color_map_vm1}. These measurements were taken in 1st $V_{max}$-increase during the cryogenic phase of the 1st cycle. For comparison, we showed the same data for Voltage map 1 as Figure \ref{color_map_vm1}.}
\label{color_map_5K_0atm}
\end{figure}

Figure \ref{color_map_vm1} indicates the color maps of 0V-subtracted data in the case of Voltage map 1 applied with various $V_{max}$ under conditions of (295 K, 1 atm), (295 K, 0 atm), and (5 K, 0 atm). Secondly, we present the color maps of 0V-subtracted data in the case of Voltage map 1, 2, and 3 applied with various $V_{max}$ under conditions of (5 K, 0 atm), as shown in Figure \ref{color_map_5K_0atm}. We can see from these results that our DM can produce wavy surface deformation corresponding to Voltage map 1, 2, and 3. The higher applied voltage causes larger displacements in all phases. These substantiate the operation of our cryogenic DM with multiple actuators. Some pixels had smaller or larger displacements than those around them regardless of temperature, which are dead or locked pixels.  

\begin{figure}[htbp]
\centering
\includegraphics[height=0.7\linewidth, angle=-90]{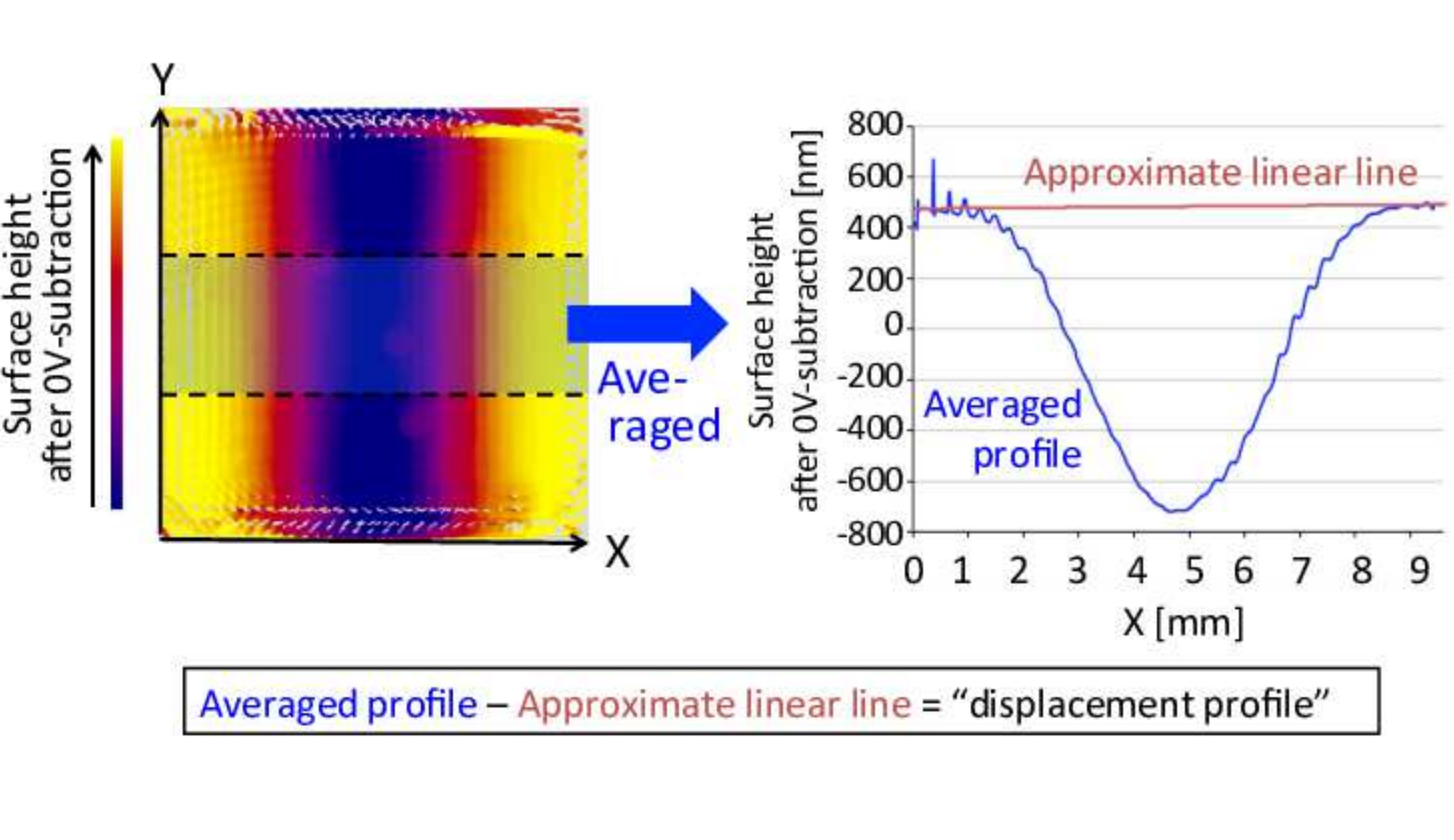}
\caption{The procedure for obtaining a displacement profile from a 0V-subtracted data.}
\label{profile}
\end{figure}

\begin{figure}[htbp]
\centering
\includegraphics[width=0.55\linewidth]{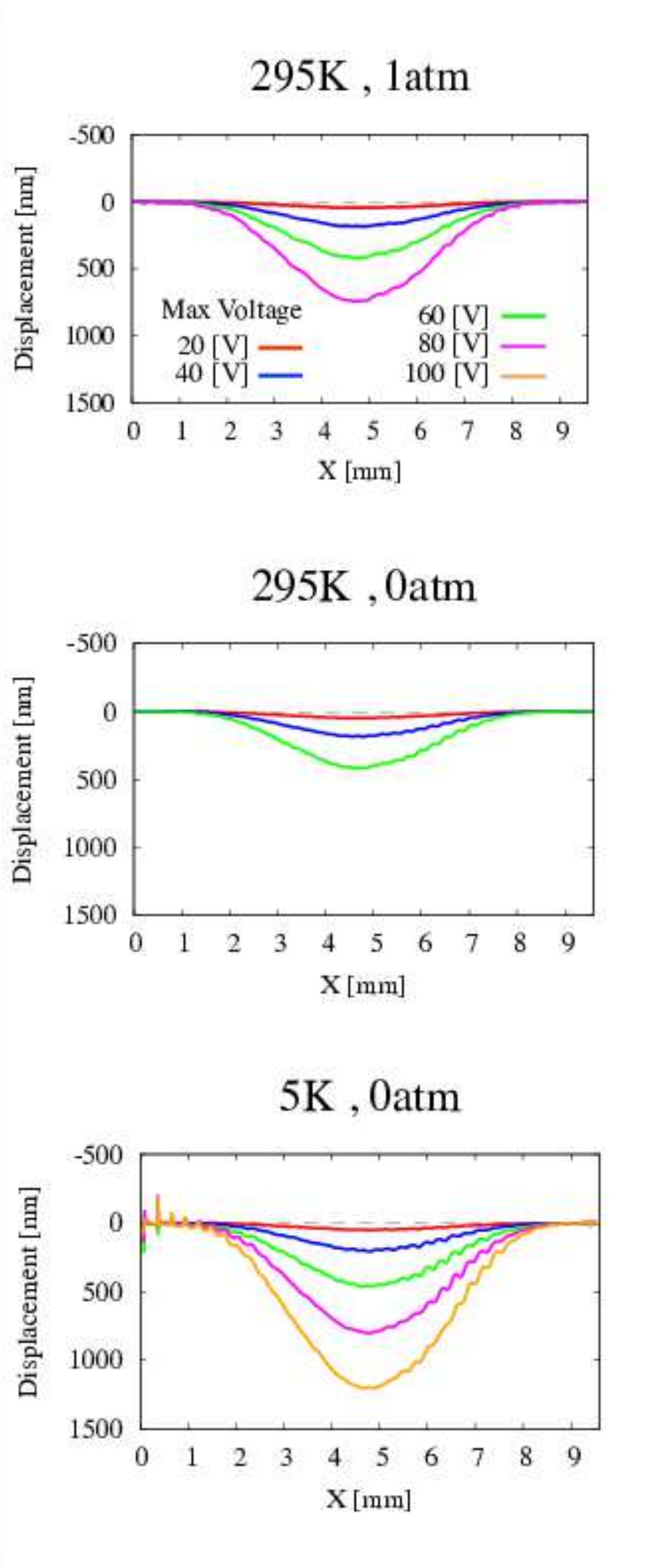}
\caption{The displacement profile of 0V-subtracted data in the case of Voltage map 1 with various values $V_{max}$ in (295 K, 1 atm), (295 K, 0 atm), and (5 K, 0 atm). These measurements were taken in the 1st $V_{max}$-increase during the initial phase and the 1st cycle.}
\label{vm1_profile}
\end{figure}

\begin{figure}[htbp]
\centering
\includegraphics[width=0.55\linewidth]{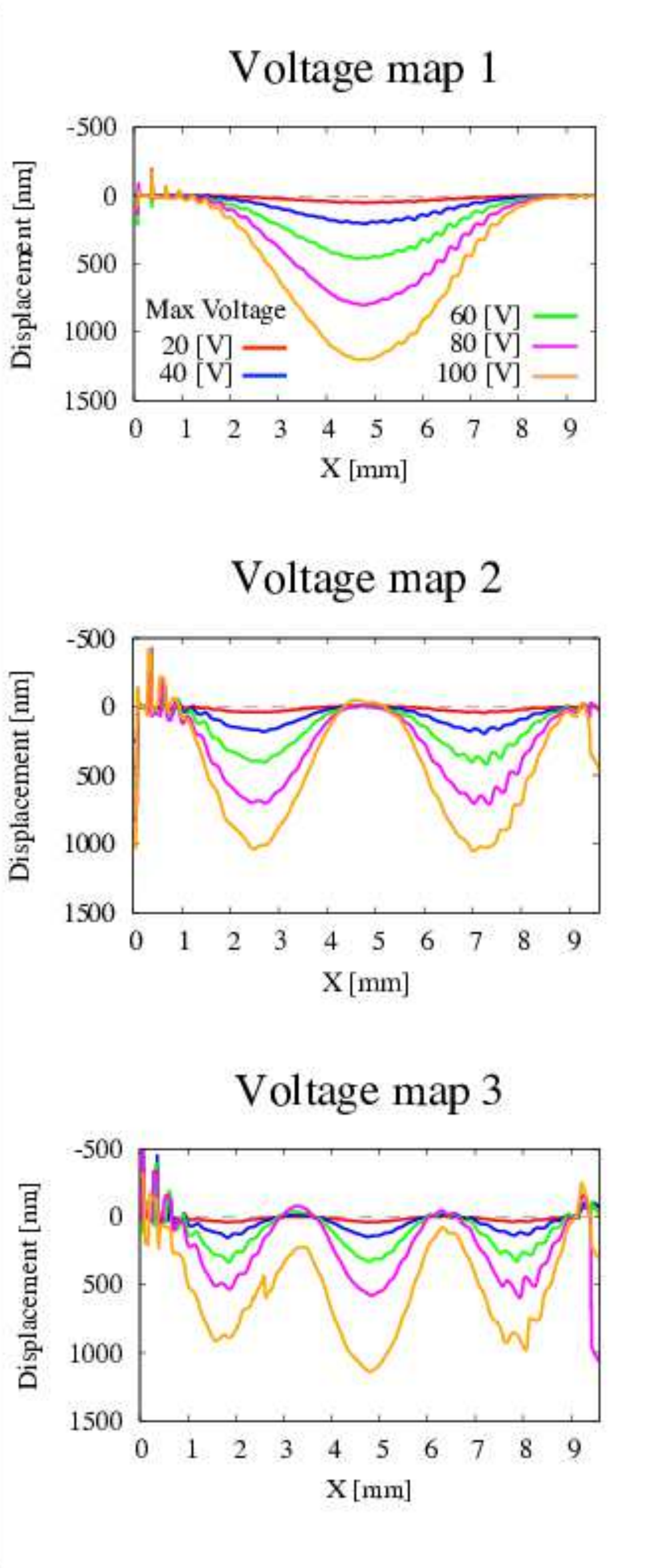}
\caption{The displacement profile of 0V-subtracted data in the case of Voltage map 1, 2, and 3 with various $V_{max}$ values in (5 K, 0 atm). The measurements were taken in the 1st $V_{max}$-increase during the cryogenic phase of the 1st cycle.}
\label{5K_0atm_profile}
\end{figure}

Next, we obtained the displacement profile of the 0V-subtracted data through the following process (see Figure \ref{profile}). First, we read out the cross-sectional profile along the X-axis averaged over the central third of $Y$ for each 0V-subtracted data. In this stage, these profiles have different offset because we measured the surface height as relative value in each surface figure. Therefore, we derived approximate linear lines using the edge region of those profiles, which is thought not to be deformed, and subtracted the profiles from each approximate linear line. Owing to the subtraction, we can also remove the slight tilt of the profile left by uncontrollably slight change of the laser-incidence angle against the DM's surface between measurements. From this process, we obtained the displacement profiles in Figure \ref{vm1_profile} and \ref{5K_0atm_profile}.

We can identify the detailed wavy structure corresponding to the size scale of a DM's actuator, which can be clearly discerned in the left-edge or rightmost slope of the profiles. The profiles at 5 K showed these structures most prominently. This may be caused by thermal deformation in each DM's actuator. 

In the edge region of the DM's surface at 5 K, we can see some discrete values in pitch of DM's actuators and they had no repeatability between measurements in other $V_{max}$-increase and decrease, while some of them became valley and some others did peaks. If we treat them as the real height of un-deformed surface and use them for the linear approximation, in some data, almost all actuators seem to have minus displacement. However, these data were 0V-subtracted and showed difference only by the voltage-applications, and therefore could not cause such whole backward deformations against the Coulomb force. For this reason, we regarded these discrete values as surface height of unexpectedly deformed points or measurement errors. Note that we therefore excluded these values from data used for the linear approximation. About Voltage map 2 and 3, the displacement becomes minus at peak region such as $x$ = 4.8 mm in the case of Voltage map 2, $x$ = 3.2, 6.4 mm in the case of Voltage map 3. These were possibly due to uncertainty of the approximate line caused by the contamination of discrete values.

In addition, the data in the case of Voltage map 3 with $V_{max}$ of 100 V jumped at the slope in the left valley and the displacement looks smaller than that in other valleys. This is because the gradient of the surface figure was too steep to be accurately measured by our system. In other supplemental tests, we confirmed that the max gradient that this system could measure was between 4.4 $\times$ 10$^{-4}$ and 7.8 $\times$ 10$^{-4}$. In this case, the surface figure has the gradient of 7.5 $\times$ 10$^{-4}$ according to the data around there and was likely to exceed the acceptable gradient range for accurate measurement.

Henceforth, we exclude some data strongly affected by unreal values from our discussion. All results of Voltage map 3 with $V_{max}$ of 100 V were removed because of both the too steep gradient and the uncertainty of the approximate line. Due to the latter reason, we also removed results of Voltage map 2 with $V_{max}$ of 100 V and with all $V_{max}$ in the 5th $V_{max}$-increase, and that of Voltage map 3 with all $V_{max}$ in the 2nd $V_{max}$-increase, with $V_{max}$ of 20, 80 V in the 4th $V_{max}$-increase, and $V_{max}$ of 60, 80 V in the 5th $V_{max}$-increase.

\subsection{Evaluation of measurement system's resolution}
\label{resolution_estimate}
\begin{figure}[htbp]
\centering
\includegraphics[width=0.55\linewidth]{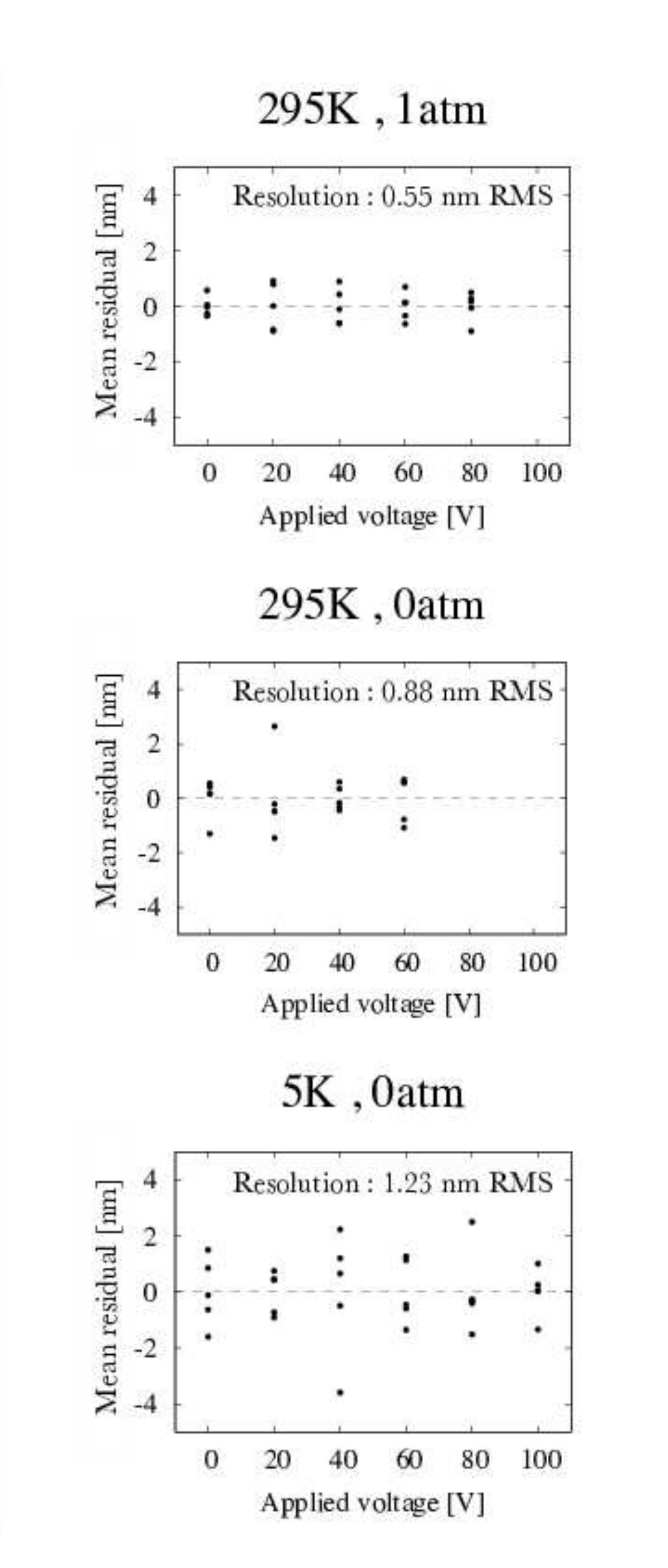}
\caption{Measurement system's resolution in (295 K, 1 atm), (295 K, 0 atm), and (5 K, 0 atm). The measurements were taken during the final phase and the 1st cycle.}
\label{resolution}
\end{figure}

We derived the displacement profiles from each result of 5 measurements of the same surface figure and read-off the maximum displacement in each profile. In this process, we used the averaged data of five measurements with no applied voltage for 0V-subtraction. The five maximum displacements were averaged at each $V_{max}$, and Figure \ref{resolution} indicates the mean residual. As we can see, they do not show strong dependence on $V_{max}$. Therefore, we evaluated the resolution of the measurement system in each phase as an RMS value of the mean residuals in all $V_{max}$. Their values in Figure \ref{resolution} were measured using only Voltage map 1. The differences between conditions can be understood if we imagine that the vacuum pump propagated vibration of the air compressor and cooling added vibration from GM-cycle coolers. The resolution at 5 K was also worsened by the uncertainty of the approximate lines used to derive maximum displacement, because those data have more discrete values in edge region than do the data at 295K. In any case, concerning Voltage map 1, we can say that the dispersion of maximum displacements due to the resolution of the measurement system is sufficiently smaller than the maximum displacements themselves in all phases.

\subsection{Evaluation of Operating Characteristics at 5 K}
We read-off the maximum displacement from each data measured in a $V_{max}$-increase and decrease. For 0V-subtraction, we subtracted the data without applied voltage measured at first in each $V_{max}$-increase and decrease. In the case of Voltage map 2 and 3, which have two or three valleys, we read the maximum displacement in each valley and averaged them. Finally, we plotted the relationship between $V_{max}$ and the maximum displacements as OC and fitted it to a quadratic function, which is assumed based on the principle of electro-static DMs~\cite{MEMS}.

\begin{figure}[htbp]
\centering
\includegraphics[width=0.6\linewidth]{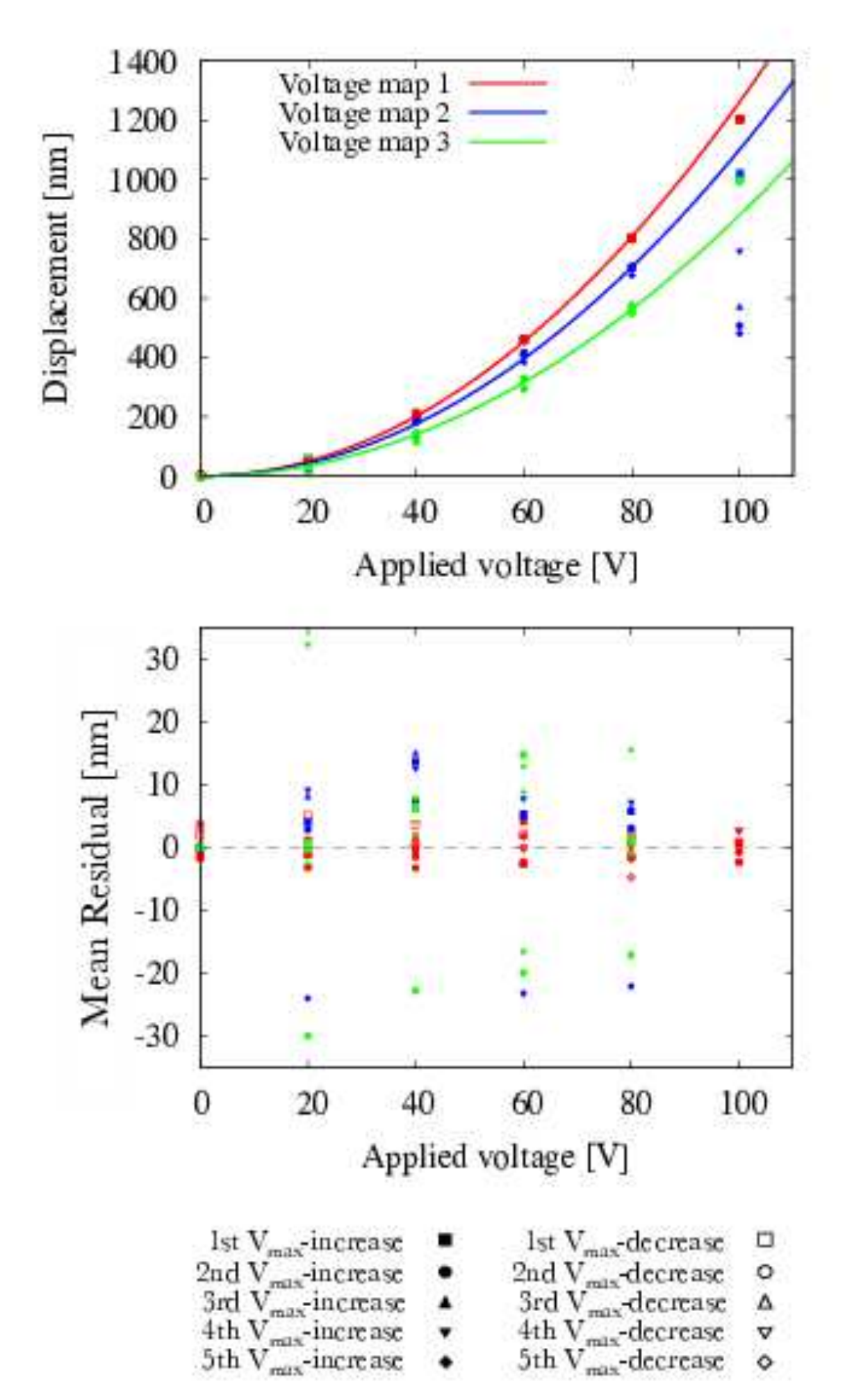}
\caption{OC at 5 K in the case of Voltage map 1, 2, and 3 (top) and their repeatabilities (bottom). Each plot indicates a measurement value and solid lines do the best-fitted curves of quadratic functions. Red, blue, and green plots and curves correspond to Voltage map 1, 2 and 3, respectively, while different shape of symbols indicates the data in different $v_{max}$-increase or decrease like legend under the graph. We did not use the data with $V_{max}$ of 100 V for the fitting, because some of them included unreal displacement (see section \ref{Result}.\ref{deformation}).}
\label{vmall_5K_0atm_OC}
\end{figure}

The top of Figure \ref{vmall_5K_0atm_OC} shows the OC plots and the fitting curve in the cryogenic phase of the 1st cooling cycle. These indicate that the OCs of our DM at 5 K are qualitatively consistent with the principle of electro-static DMs. The coefficient of the quadratic term is determined by the physical properties and the microscopic geometry of the chip. Because we can not disclose proprietary information about BMC products, the quantitative discussion is withheld from this paper.

\subsection{Evaluation of operating hysteresis and repeatability at 5 K}
\label{repeatability_estimate}
To evaluate operating repeatability at 5 K, we compared maximum displacements of the same $V_{max}$ in 1st to 5th $V_{max}$-increase and decrease. These are averaged, and the mean residuals were plotted at the bottom of Figure \ref{vmall_5K_0atm_OC}. The dispersion seems to be independent of $V_{max}$. Therefore, we evaluated the operating repeatability as the RMS value of the mean residuals at all $V_{max}$. The derived repeatabilities at 5 K were 2.10, 2.16, and 3.53 nm RMS for Voltage map 1, 2, and 3, respectively. Though these values affect the dispersion of maximum displacements slightly more than the measuring system's resolution does, they are still sufficiently smaller than the maximum displacements themselves. In addition, no hysteresis was detected since we did not find more significant differences in OCs between increase and decrease of $V_{max}$ compared with repeatability.

\subsection{Comparison of Operating Characteristics}

\subsubsection{Difference due to spatial frequency}
\label{spacial_frequency}
We compared OCs in the case of Voltage map 1, 2, and 3 at 5 K in Figure \ref{vmall_5K_0atm_OC}. It is clear that the maximum displacements at the same $V_{max}$ become smaller, as the spatial frequency of the Voltage map becomes larger. The character of the membrane surface can explain this. Tensile force in the membrane surface is proportional to the height gradient between neighboring actuators and the elastic coefficient. The height gradient becomes larger in surface figures of higher spatial frequency. Therefore, the effects of tensile forces can be larger and displacements can be more suppressed in Voltage map of higher spatial frequency.

\subsubsection{Difference due to temperature}
\begin{figure}[htbp]
\centering
\includegraphics[height=0.5\linewidth]{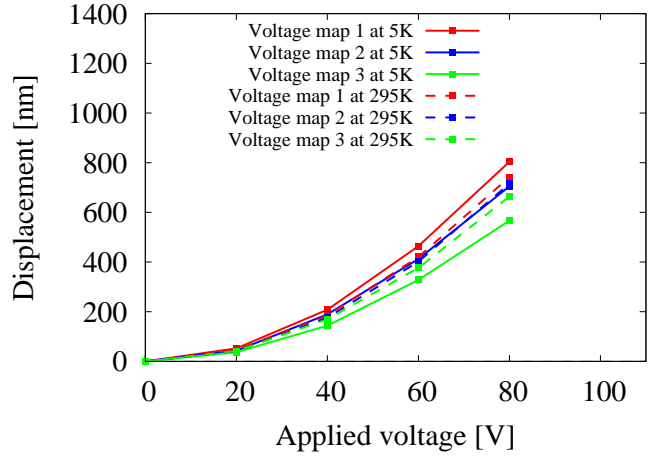}
\caption{OCs at (295 K, 1 atm) in the initial phase and (5 K, 0 atm) in the 1st cycle in the case of Voltage map 1, 2, and 3. The measurements were taken in the 1st $V_{max}$-increase in each phase.}
\label{vmall_temp_compare}
\end{figure}

We compared OCs between the conditions of (295 K, 1 atm) and (5 K, 0 atm) in Figure \ref{vmall_temp_compare}. In the case of Voltage map 1, OC at 5 K showed significantly larger maximum displacements than those of the same $V_{max}$ at 295 K. This may be because the electro-statically actuated diaphragm approached the electrode by thermal construction of the DM chip, and the effect of Coulomb force became larger even if we applied the same voltages. However, maximum displacements under the same $V_{max}$ were almost identical between 295 K and 5 K in the case of Voltage map 2, and in the case of Voltage map 3, those at 295 K were larger than those at 5 K. One possible reason may be that the tensile force of the membrane surface became larger at lower temperature because the elastic coefficient of silicon becomes large there~\cite{Si}. Since tensile force is more effective in a Voltage map of higher spatial frequency (see section \ref{spacial_frequency}), displacements became smaller at lower temperatures in the case of Voltage map 3, even given the effect of thermal construction.

\subsubsection{Difference due to air pressure}
When we compared OCs between the conditions of (295 K, 1 atm) and (295 K, 0 atm), they showed little difference as in Figure \ref{vm1_OC_cycle_compare}. It can be said that air pressure had no detectable effect on OC in this cooling test.

\subsubsection{Difference due to cooling cycles}
\begin{figure}[htbp]
\centering
\includegraphics[width=0.5\linewidth]{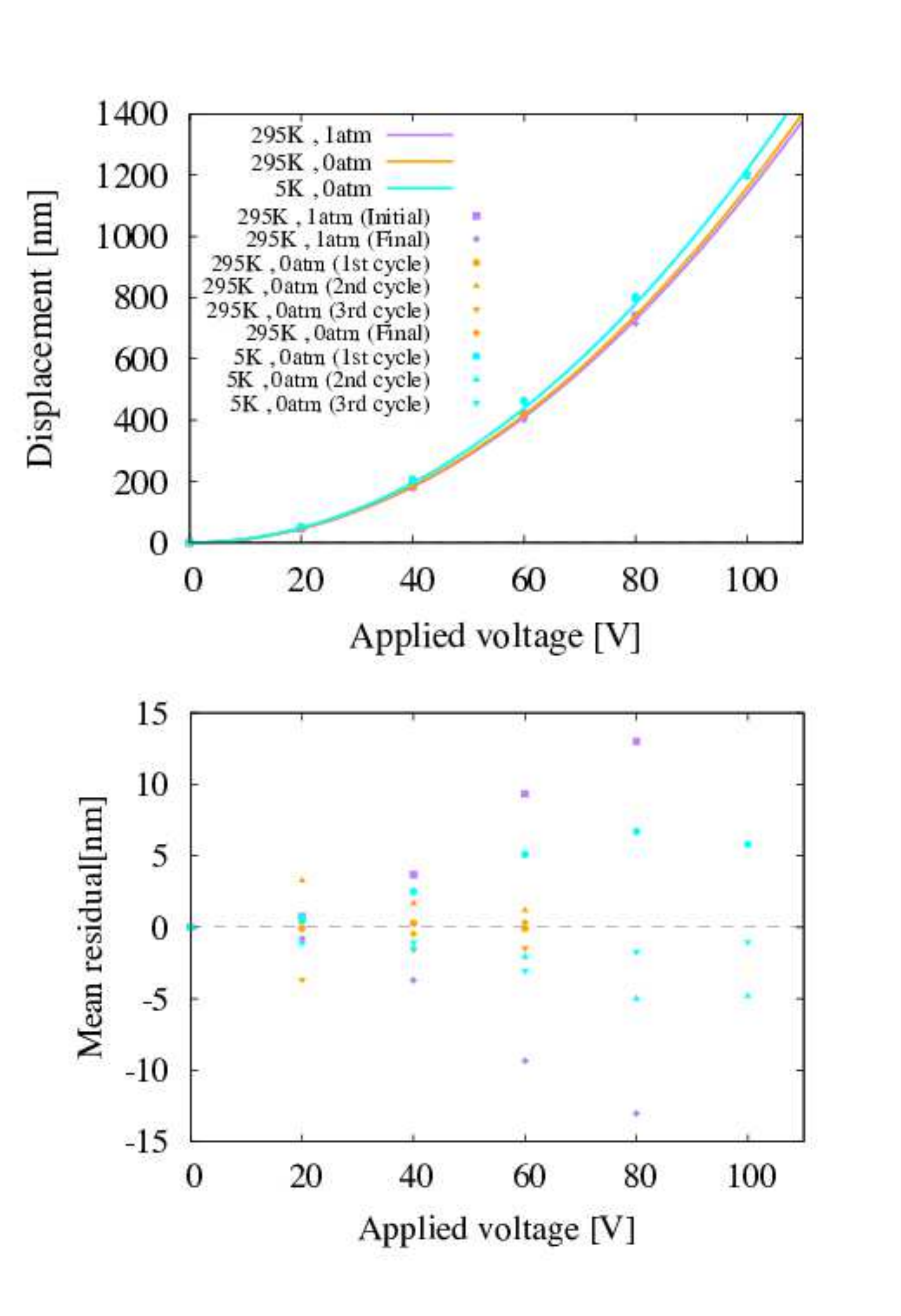}
\caption{OCs in each phase of three cooling cycles in the case of Voltage map 1 (top) and comparison at the same temperature and air pressure (bottom).}
\label{vm1_OC_cycle_compare}
\end{figure}

Figure \ref{vm1_OC_cycle_compare} shows the change in OCs during cooling cycles. When we focus on the OCs in (5 K, 0 atm), mean residuals in all cooling cycles are at almost the same level as dispersion due to repeatability, and no significant difference is seen in the OCs of three cooling cycles. This can also be said about OCs in (295 K, 0 atm). Although the OC in the final phase of (295 K, 1 atm) has smaller maximum displacements at the same $V_{max}$ than that in the initial phase of (295 K, 1 atm), our DM's operation is at least durable against repeated cooling and warming in a vacuum environment.

\clearpage
\section{Discussion}

\subsection{Correction of convexity}
As mentioned in section \ref{Result}.\ref{no applied-voltages}, the surface figure with no applied voltage has convexity at 5 K. There are some approaches we can take to correct wavefront error using a DM with such a surface figure.

One approach is to make a required surface figure for wavefront correction only by applying voltages starting from the convex surface. For our DM, however, the convexity's PV value of a few $\mu$m exceeds the stroke of our DM. Therefore, we have to roughly correct the convexity in other approaches and correct only the left convexity using the actuator's stroke.

\begin{figure}[htbp]
\centering
\includegraphics[height=0.7\linewidth,angle=90]{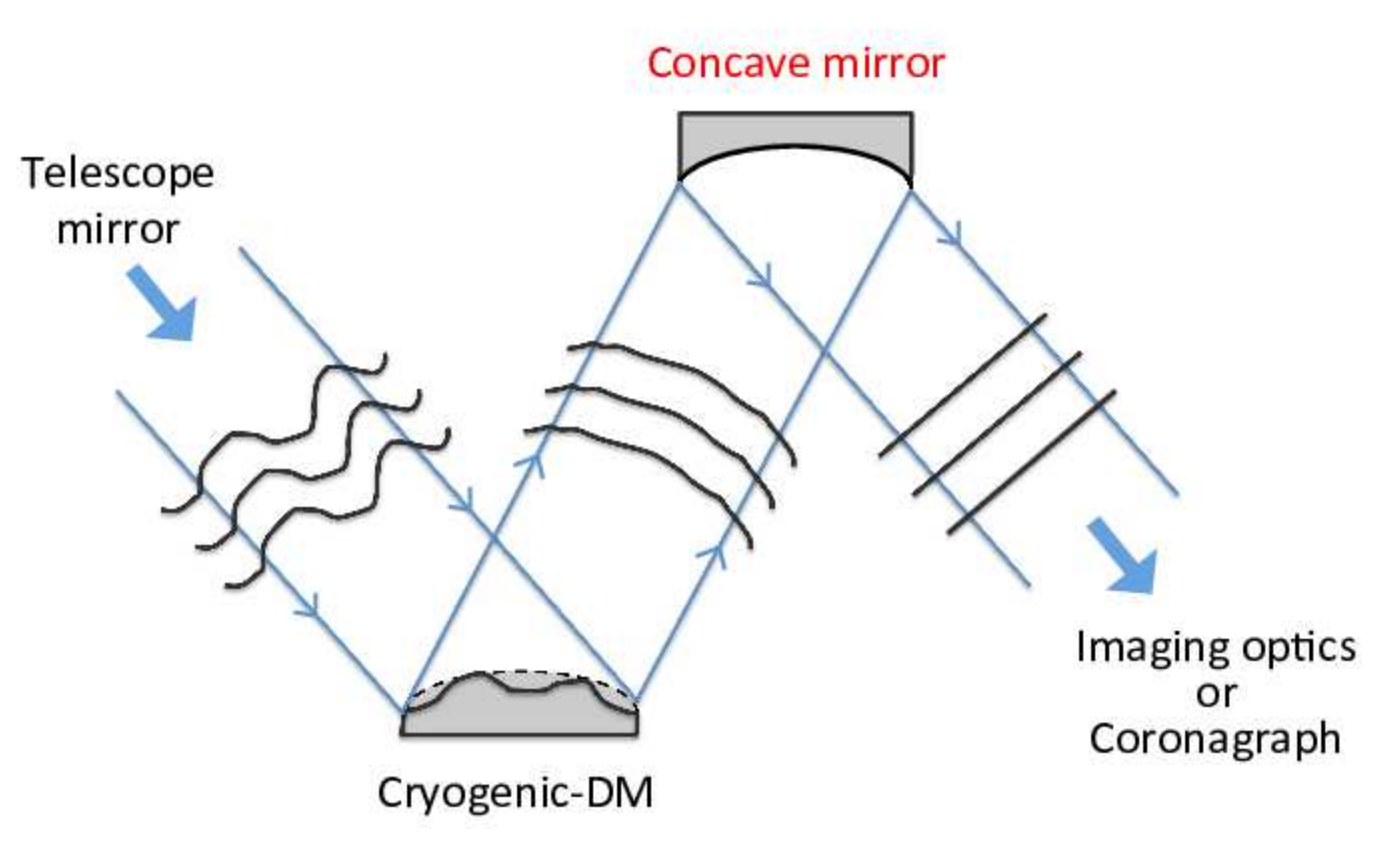}
\caption{Correction of convexity by fixed concaved mirror}
\label{correct_mirror}
\end{figure}

For the rough correction, for example, we can cancel the convexity by reflection in another fixed mirror after reflection in our DM, as in Figure \ref{correct_mirror}. This fixed mirror must have a concave surface, inverse to the DM's surface.

In addition, methods to make the DM's surface roughly flat at 5 K are worth to study. As one possibility, we can hold the AlN board to apply stress so that the DM chip is concave at ambient temperature and becomes flat when cooled to 5 K. It also can be possible to glue the center of the DM chip to AlN board.

\subsection{Accuracy of wavefront correction}
Space-borne telescopes do not need DMs to realize the required surface figure when applying voltages only once, because they are not affected by perturbations in the air and experience no quick changes of wavefront error. Therefore, we can finely control the applied voltage to each actuator after roughly realizing the required surface figure. However, the final accuracy with which the required surface figure is realized is limited by the following two factors:

First, the control resolution of an actuator must be considered. The controllable minimum unit of the applied voltages was about 4 mV in DM's control system used in this test (see section \ref{Cooling tests}.\ref{control_system}). The corresponding displacement resolution, $\Delta Z$, is written as the following equation. 
\begin{equation}
\Delta Z = a(V + \Delta V)^{2} - aV^{2} \simeq 2aV \Delta V
\end{equation}
In this equation, $a$ shows the coefficient when we fit the OCs to $Z = a V^{2}$, while displacements and applied voltages are indicated by $Z$ and $V$. According to this test, our results at 5 K are $a$ = 0.12, 0.11, and 0.09 for Voltage map 1, 2, and 3, respectively. Even considering the maximum applied voltage in this test, 100 V, the displacement resolutions will be $9.6\times10^{-2}$ nm, $8.8\times10^{-2}$ nm, and $7.2\times10^{-2}$ nm for Voltage map 1, 2, and 3, respectively.

As a second reason for limiting the final accuracy, we note the operational repeatability of the DM. Our test resulted in the repeatability at 5 K of 2.6 nm RMS averaged in Voltage map 1, 2, and 3. This value is much larger than the control resolutions discussed previously. The operating repeatability obtained in this test seems to be more dominant as a reason for limiting the accuracy compared to the control resolutions.

Now, we take 2.6 nm RMS as the final accuracy with which the surface figure required for wavefront correction can be realized. When light is reflected by the DM's surface, optical path difference becomes double of the difference in the DM's surface height. As a result, we can assume that the reflected light after wavefront correction has a wavefront error of 5.2 nm RMS.

Here, It is noted that the resulting surface figure at 5 K included discrete values in the edge region. Even if we avoid these values and use 26 $\times$ 26 actuators in the center of the DM's surface, we can correct wavefront errors with spatial frequencies up to 13 cycle/D [m$^{-1}$].

\subsection{Simulation of a wavefront correction}

\subsubsection{Wavefront after correction}
An actual wavefront error is comprised of a mix of components with various spatial frequencies. Power spectral density (PSD) indicates the contribution of each components as a function of spatial frequency. For telescopes, PSDs are empirically written as Lorenz functions of spatial frequency, $\rho$, as follows~\cite{PSD}:
\begin{align}
PSD(\rho) = \frac{\sigma_{0}^{2} A}{h_{0}} \frac{1}{1+ \left( \dfrac{\rho}{\rho_{c}} \right) ^{a}} \label{eqPSD}\\
h_{0} = \iint_{aperture} \frac{1}{1+ \left( \dfrac{\rho}{\rho_{c}} \right) ^{a}} d \zeta d \eta
\end{align}
The index $a$ and half width at half maximum of Lorenz function, $\rho_{c}$ [m$^{-1}$], are constants determined by optics. $\sigma_{0}$ [nm RMS] indicates the RMS value of the wavefront error and A [m$^2$] indicates the aperture area. We set the 2D position coordinate $(\zeta, \eta)$ in the pupil plane. For example, the Hubble Space Telescope with aperture diameter of 2.4 m gives $a$ = 2.9 and $\rho_{c}$ = 4.3 m$^{-1}$, and the Very Large Telescope with aperture diameter of 8.2 m gives $a$ = 3.1 and $\rho_{c}$ = 0.35 m$^{-1}$~\cite{HST_and_VLT}.

We simulated the case of the next-generation Japanese infrared satellite, SPICA. In 2012, the SPICA Coronagraph Instrument was considered to be loaded onto the spacecraft. SPICA was assumed to have an aperture with diameter of 3 m and the wavefront error with PSD given by equation (\ref{eqPSD}) at $a$ = 3.0, ~$\rho_{c}$ = 1.0 m$^{-1}$, and $\sigma_{0}$ = 350 nm RMS. Based on these values, we can achieve the diffraction limit at 5 $\mu$m with a Strehl ratio larger than 80 $\%$. 

\begin{figure}[htbp]
\centering
\includegraphics[height=0.7\linewidth, angle=-90]{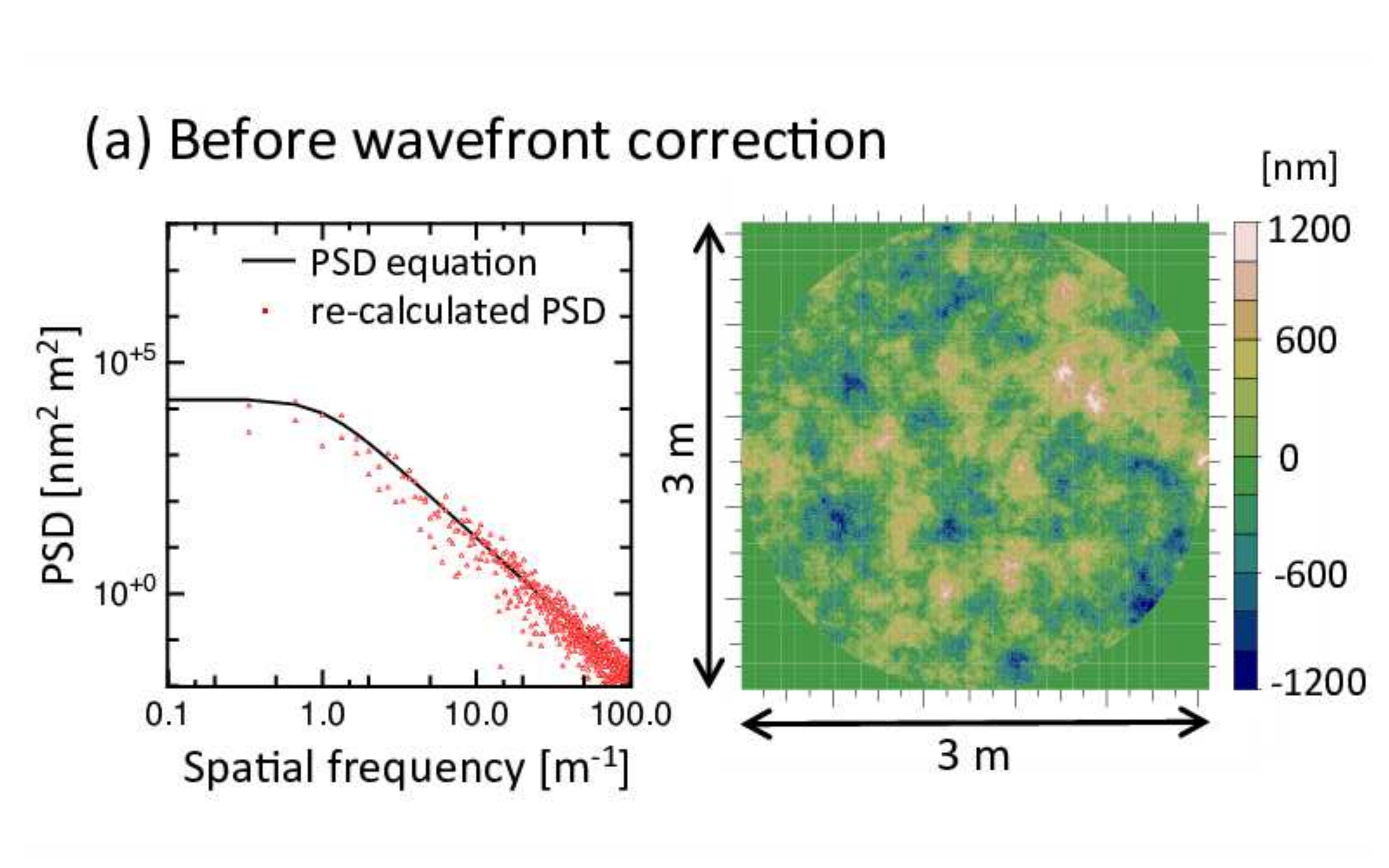}\\
\includegraphics[height=0.7\linewidth, angle=-90]{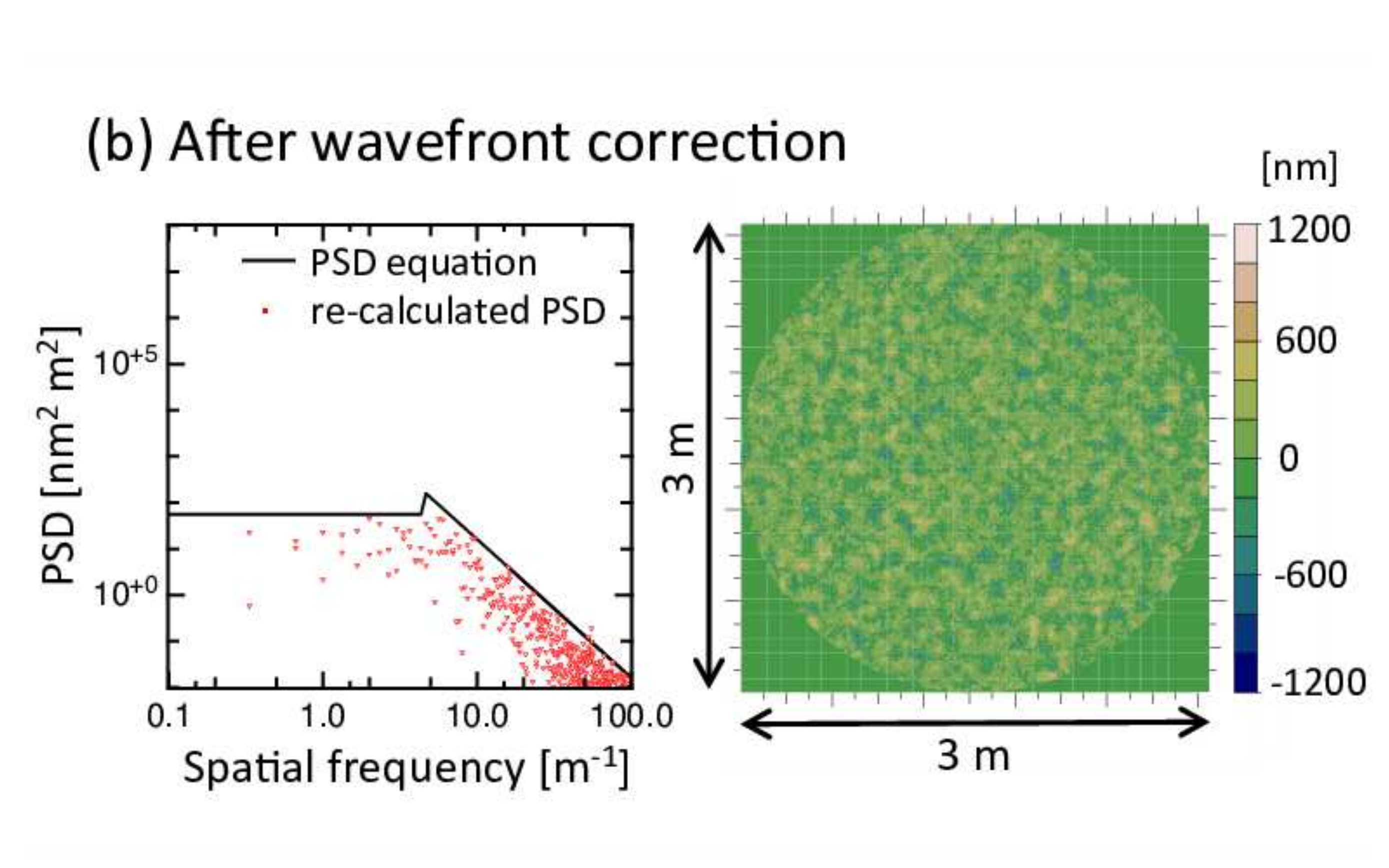}
\caption{PSD of a wavefront error (left) and an example of the wavefront error's spatial distribution in the pupil plane (right), (a) Before wavefront correction (like equation (\ref{eqPSD})) and (b) After wavefront correction (like equation (\ref{eqPSD_2})). We assumed a wavefront error of $\sigma$ = 350 nm RMS before the correction. In graphs, black solid lines indicate the PSD expressed by equations and red points indicate PSDs re-calculated from the right spatial distributions.}
\label{PSD}
\end{figure}

Against such optics, we assumed that our DM could improve the wavefront error with lower spatial frequency than 13 cycle/D [m$^{-1}$] to 5.2 nm RMS. In this case, we could realize the following PSD owing to wavefront correction by our DM at $A$ = $\pi \times (3/2)^{2}$ m$^2$, ~$a$ = 3.0, ~$\rho_{c}$ = 1.0 m$^{-1}$, ~$\sigma_{0}$ = 350 nm RMS, ~$\sigma_{1}$ = 5.2 nm RMS, and $h_{1}$ = 13/3:
\begin{equation}
PSD(\rho) = \begin{cases}
               \dfrac{\sigma_{1}^{2} A}{h_{1}} & ( 0 \leq \rho \leq 13/3 ) \\
               \\
               \dfrac{\sigma_{0}^{2} A}{h_{0}} \dfrac{1}{1+ \left( \dfrac{\rho}{\rho_{c}} \right) ^{a}} & ( 13/3 < \rho )
            \end{cases}
            \label{eqPSD_2}
\end{equation}
At the left side of Figure \ref{PSD}, we show the PSD profile before and after wavefront correction. From the PSD, we simulated an exemplary spatial distribution of the wavefront error in the pupil plane, as shown at the right side of Figure \ref{PSD}. The wavefront error was suppressed significantly in the corrected region of spatial frequency.

\subsubsection{PSF with a circular aperture}
\begin{figure}[htbp]
\centering
\includegraphics[height=0.7\linewidth, angle=-90]{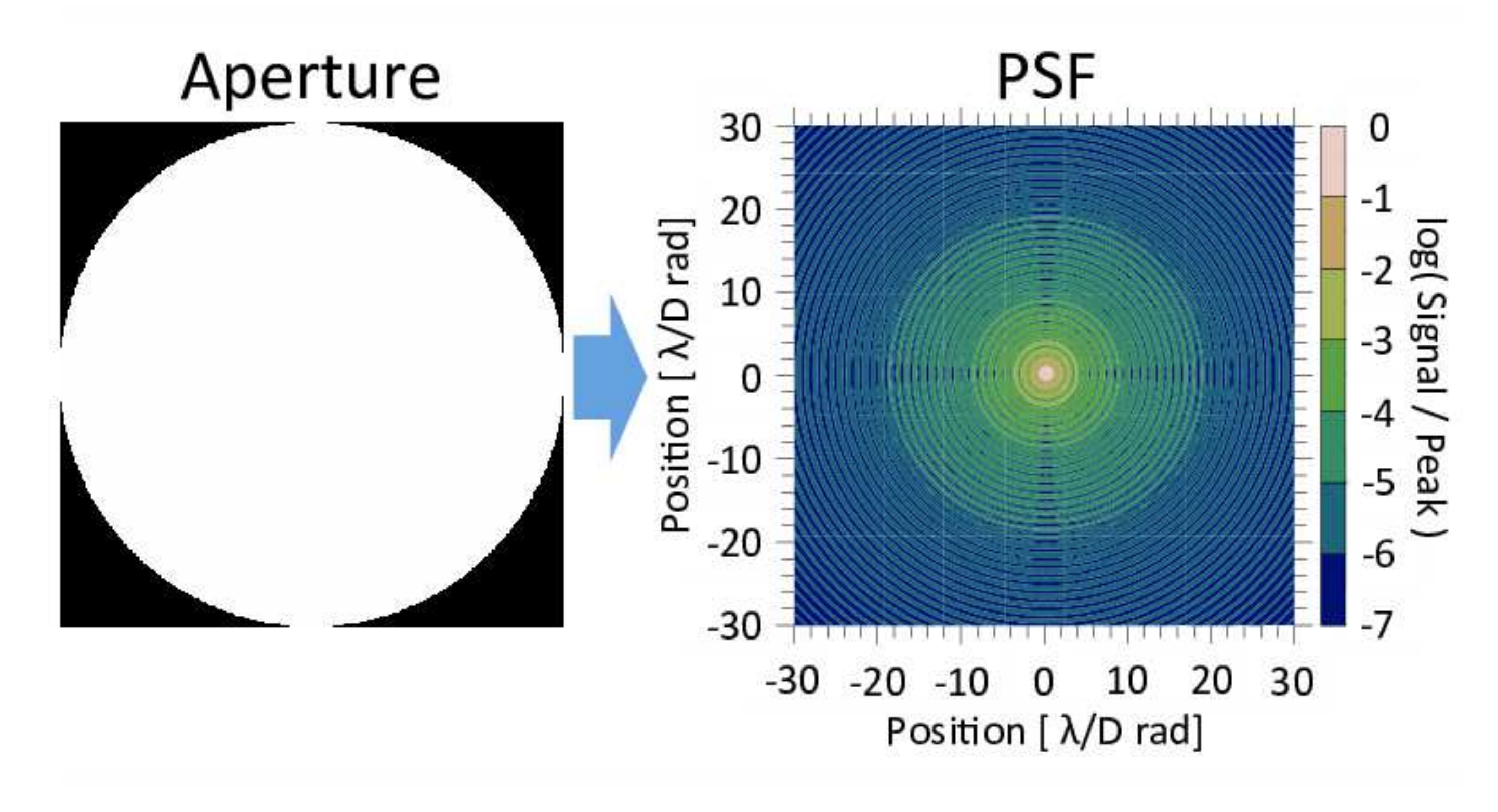}
\caption{Circular aperture (black and white region have transmittances of 0 $\%$ and 100 $\%$, respectively) and the PSF simulated with no wavefront error at 1 $\mu$m.}
\label{Airy_mask_PSF}
\end{figure}

\begin{figure}[htbp]
\centering
\includegraphics[height=0.7\linewidth, angle=-90]{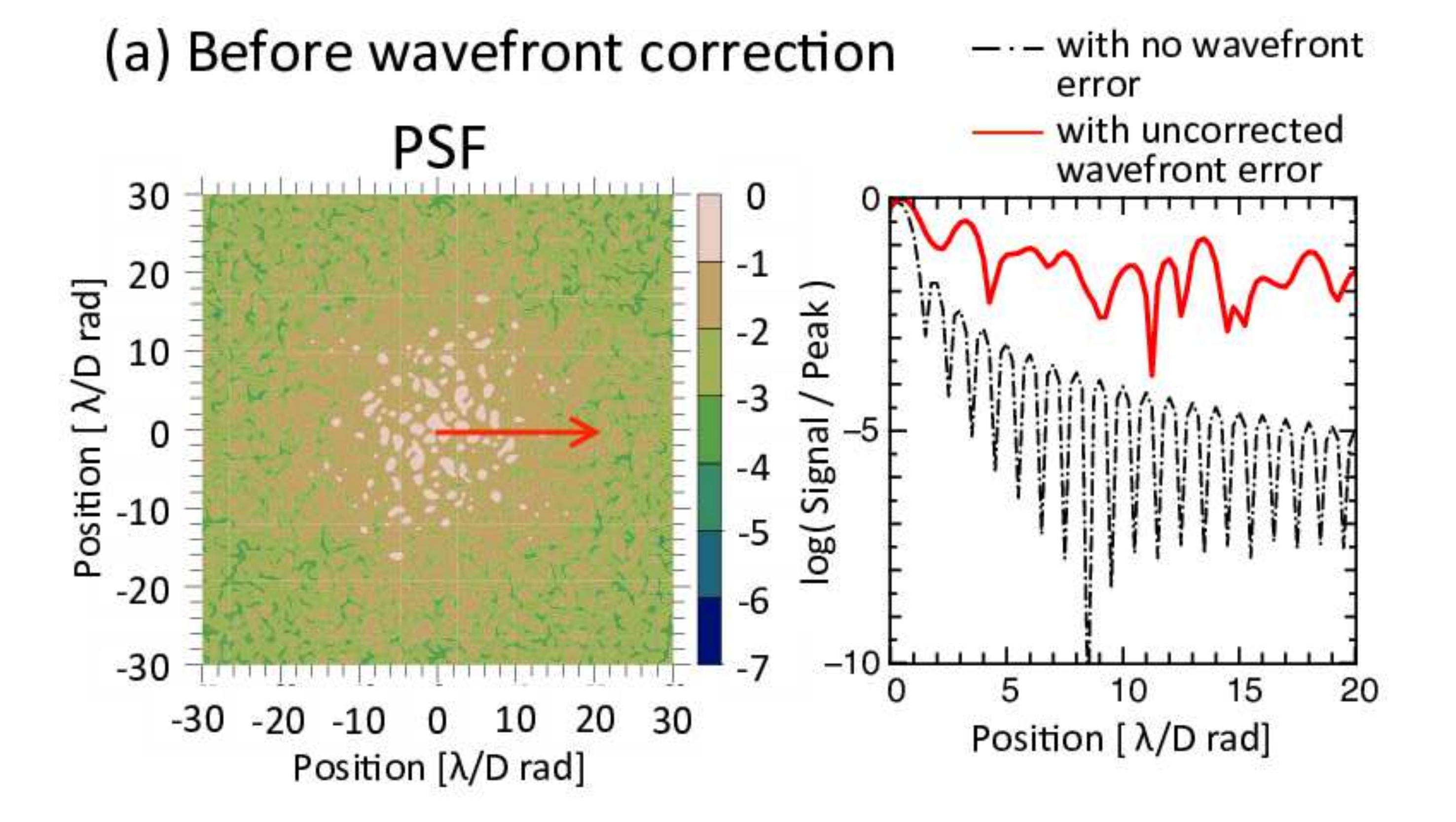}\\
\includegraphics[height=0.7\linewidth, angle=-90]{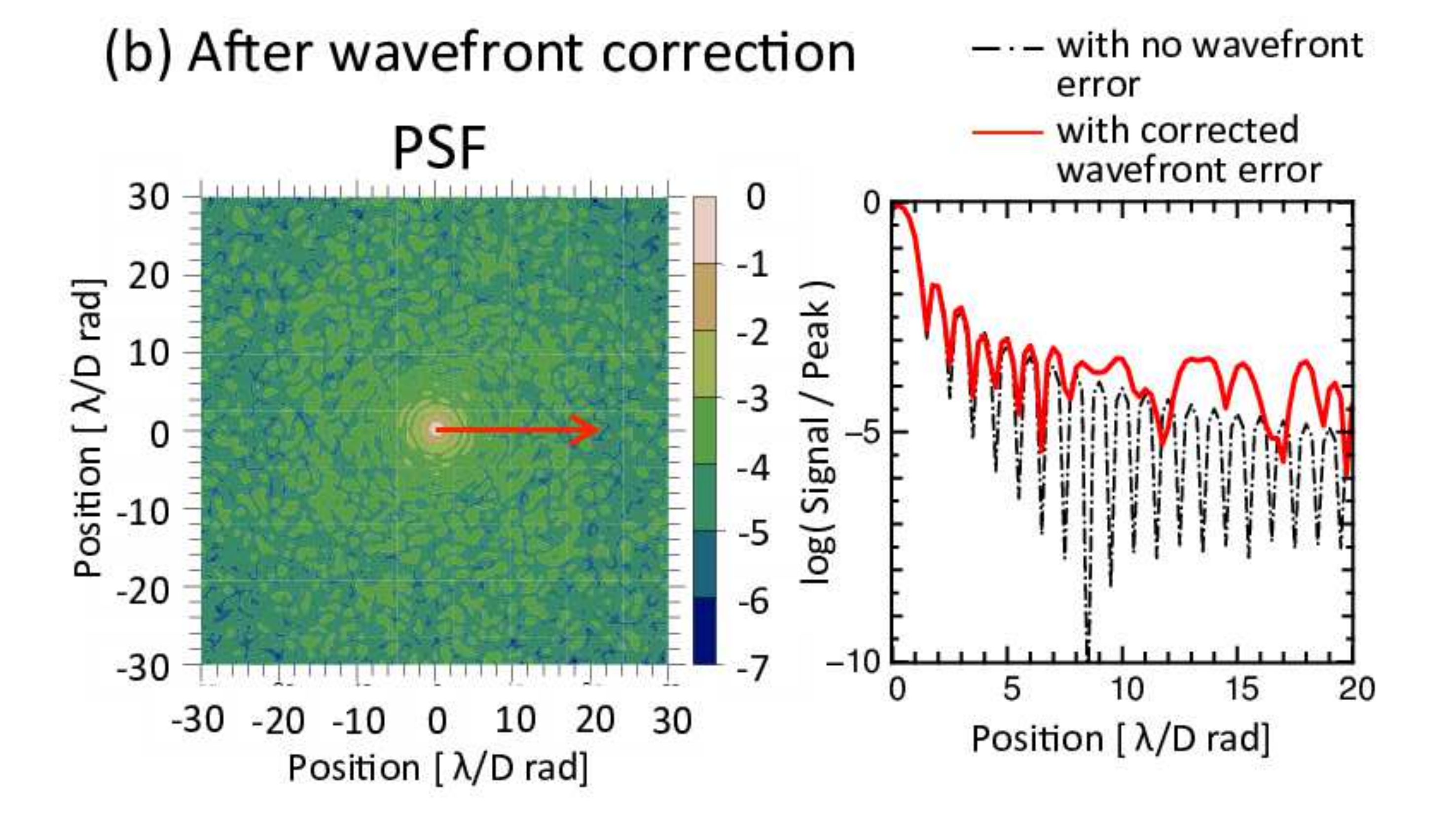}
\caption{Simulated PSF at 1 $\mu$m from the circular aperture, (a) before wavefront correction and (b) after wavefront correction. We assumed wavefront errors of $\sigma$ = 350 nm RMS before the correction. Since these PSFs are normalized by each peak value, they indicate the contrast to the peak value. The plots on the right show the cut profiles along the red arrows. The red solid lines mean PSF profile with wavefront error and black dashed lines mean that with no wavefront error.}
\label{Airy_PSF_WFC}
\end{figure}

We simulated a PSF example using a Fourier transformation in the case of a pupil function with a circular aperture like that in Figure \ref{Airy_mask_PSF}. In this section, we present that the wavefront correction using our DM enabled diffraction limit at shorter wavelength than the designed value even in the same telescope. If we observe at 1 $\mu$m using a telescope with a circular aperture of 3 m diameter and the same wavefront error assumed for SPICA in 2012 (like equation (\ref{eqPSD_2})), we obtain a PSF like Figure \ref{Airy_PSF_WFC}. By adding our DM to the optics, we can not only realize the PSF without wavefront error in the region < 5 $\lambda/D$ [rad], but also improve contrast in the outer region, even at 1 $\mu$m. This means that we can obtain images almost with a diffraction limit of 1 $\mu$m using a telescope designed to achieve a diffraction limit of 5 $\mu$m. Thus, wavefront correction by our DM will make it possible to observe in a diffraction limit at shorter wavelength using telescopes with the same surface accuracy.

In addition, we can relax our requirement against surface figure error of space-borne infrared telescopes. This will contribute to achieving such telescopes with the same optical performance in lower cost and risk and over shorter timescales.

\subsubsection{PSF with a coronagraph aperture}
\begin{figure}[htbp]
\centering
\includegraphics[height=0.7\linewidth, angle=-90]{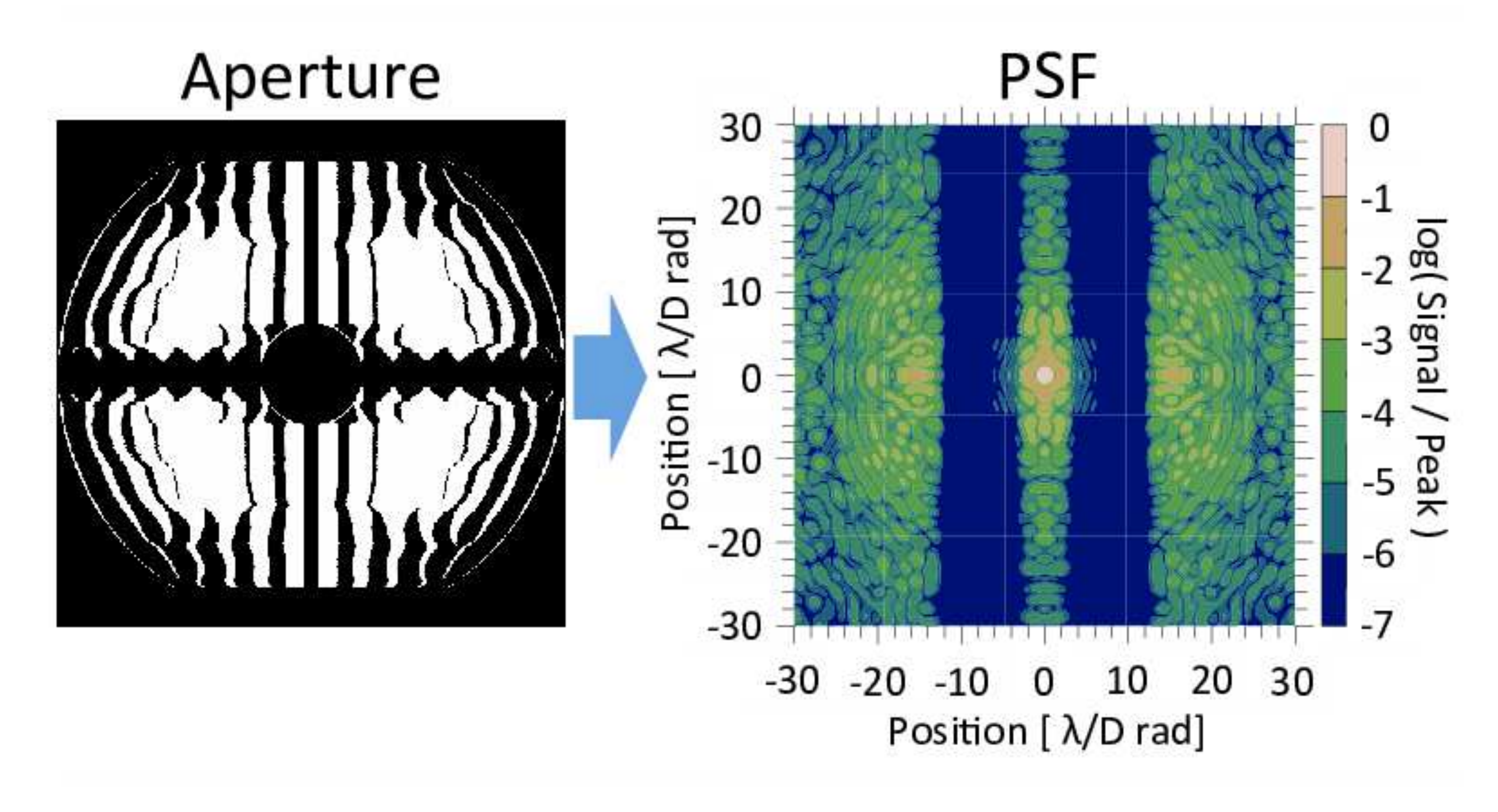}
\caption{Aperture of a binary-shaped pupil-mask coronagraph used in this simulation (black and white regions have transmittances of 0 $\%$ and 100 $\%$, respectively) and the PSF is simulated with no wavefront error at 5 $\mu$m.}
\label{coronagraph_mask_PSF}
\end{figure}

\begin{figure}[htbp]
\centering
\includegraphics[height=0.7\linewidth, angle=-90]{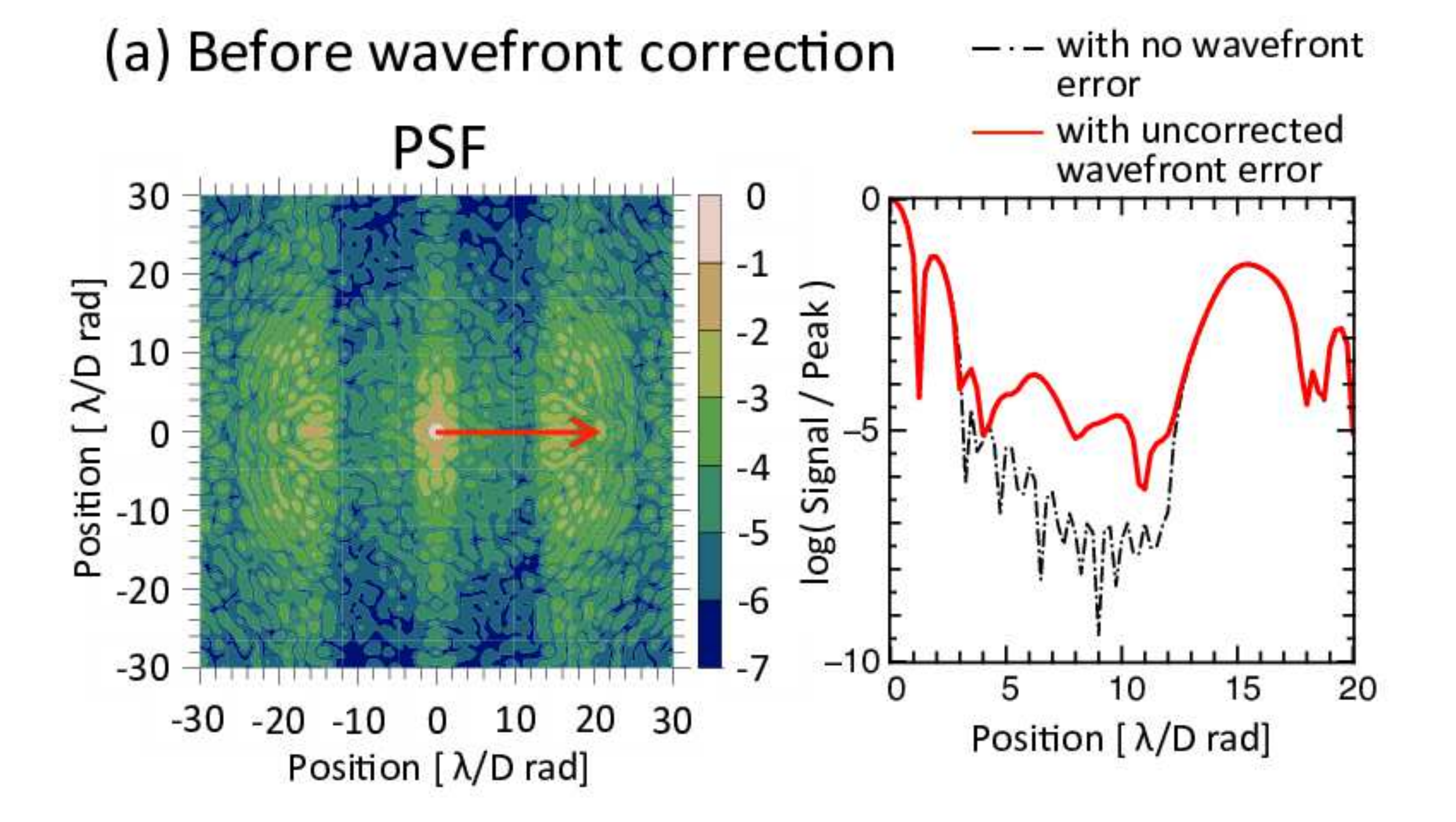}\\
\includegraphics[height=0.7\linewidth, angle=-90]{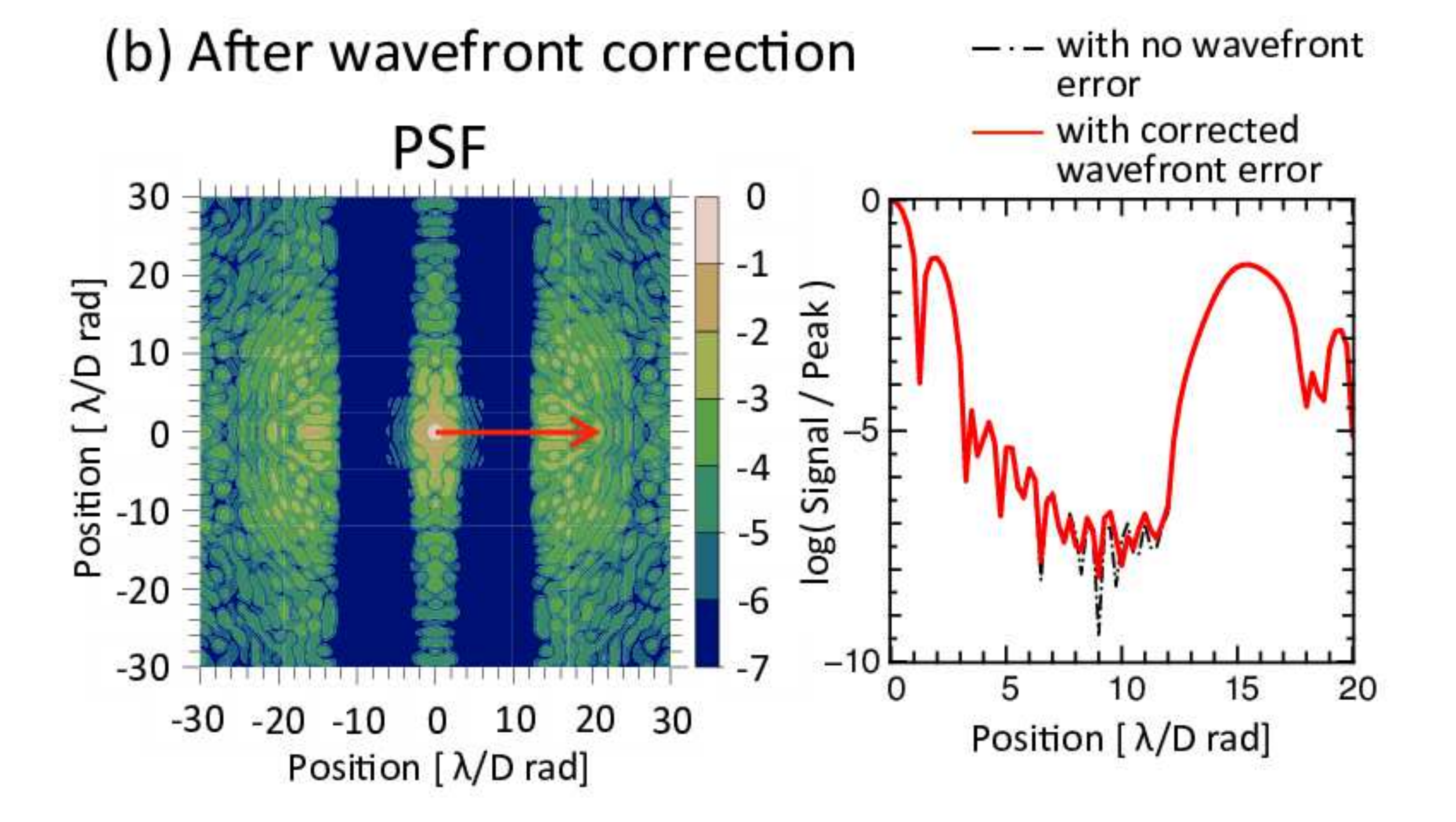}
\caption{Simulated PSF at 5 $\mu$m from the coronagraph aperture, (a) before wavefront correction and (b) after wavefront correction. Detail explanation is same as Figure \ref{Airy_PSF_WFC}.}
\label{coronagraph_PSF_WFC}
\end{figure}

Next, we simulated the PSF in the case where we used our DM with coronagraph optics. As an example, we assumed a binary-shaped pupil-mask coronagraph, which is capable of controlling the PSF with a mask like Figure \ref{coronagraph_mask_PSF} on the pupil plane. Figure \ref{coronagraph_PSF_WFC} shows the comparison between PSFs at 5 $\mu$m before and after wavefront correction. The contrast of 10$^{-5}$ - 10$^{-4}$ in dark region ( 5 - 12 $\lambda/D$ [rad] ) is improved to 10$^{-7}$ - 10$^{-6}$ with the addition of our DM. Contrast after wavefront correction can be said to have became significantly close to the theoretical value without wavefront error. Note that observation at longer wavelengths is less sensitive to a wavefront error, because a PSF is the Fourier transformation of the function with $\phi(\zeta, \eta) / \lambda$, when $\phi$ means the wavefront error's amplitude at each position in pupil plane ($\zeta, \eta$) and $\lambda$ is the observation wavelength. Therefore, higher contrast is expected at longer wavelengths than 5 $\mu$m.

If the PSF of a central star closer than 10 pc from the sun has such a dark region, we can directly observe the planets rotating with a semi-major axis of a few 10s AU. Such high-contrast in mid-infrared observation enables the first detection of dark planets with old ages of 1 - 5 Gyr and light masses of a few $M_{J}$. This will contribute to revealing the diversity of planetary systems and understanding the universal process of planetary formation and evolution.

\section{Summary}
We aim to correct a wavefront error in space-borne infrared telescopes on orbit using small cryogenic DMs. For this purpose, we developed a new MEMS-processed electro-static DM with 1,020 actuators usable at 5 K. It was successfully operated at 5 K and the OC was qualitatively consistent with the principle of electro-static DMs. We found no hysteresis and the operating repeatability of 2.6 nm RMS in the case of Voltage map 1, 2, and 3. In addition, it remained durable over three cooling cycles. Such a DM enables us to make observations in diffraction limits at shorter wavelengths than the designed value. This contributes to the development of space-borne infrared telescopes with lower cost and risk and over shorter time scale than ever. In addition, if we use our DM in combination with coronagraph optics in space-borne infrared telescopes, we can directly observe darker planets than ever before. This helps us to generally understand the formation of planetary systems.

\vspace*{2.0\baselineskip}
We are grateful to T. Mita, T. Wada, and T. nakagawa for his kind support. The opportunity to use the interferometer was provided by JAXA. This work was supported by KAKENHI, Grant-in-Aid for Scientific Research (A), 22244016. We sincerely thank for fruitful reviewer's comments and editorial works.

\end{document}